\documentclass[10pt,twocolumn,letterpaper]{article}

\usepackage[pagenumbers]{cvpr} 

\usepackage[dvipsnames]{xcolor}

\definecolor{cvprblue}{rgb}{0.21,0.49,0.74}
\usepackage[pagebackref,breaklinks,colorlinks,citecolor=cvprblue]{hyperref}

\usepackage{multirow}
\usepackage{amssymb}
\usepackage{colortbl}  
\usepackage{xcolor}
\usepackage{framed}
\usepackage{color}
\usepackage{array}   
\usepackage{algorithm}
\usepackage{algorithmic}
\usepackage[normalem]{ulem}
\useunder{\uline}{\ul}{}
\usepackage{balance}

\def\paperID{4775} 
\def\confName{CVPR}
\def\confYear{2024}

\title{LFSRDiff: Light Field Image Super-Resolution via Diffusion Models}

\author{
 Wentao Chao\textsuperscript{\rm 1},
 Fuqing Duan\textsuperscript{\rm 1*},
 Xuechun Wang\textsuperscript{\rm 1}, 
 Yingqian Wang\textsuperscript{\rm 2},
 Guanghui Wang\textsuperscript{\rm 3}
\\
\textsuperscript{\rm 1}Beijing Normal University, \textsuperscript{\rm 2} National University of Defense Technology\\
\textsuperscript{\rm 3} Toronto Metropolitan University \\
}

\begin{document}
\maketitle
\begin{abstract}
Light field (LF) image super-resolution (SR) is a challenging problem due to its inherent ill-posed nature, where a single low-resolution (LR) input LF image can correspond to multiple potential super-resolved outcomes. Despite this complexity, mainstream LF image SR methods typically adopt a deterministic approach, generating only a single output supervised by pixel-wise loss functions. This tendency often results in blurry and unrealistic results. Although diffusion models can capture the distribution of potential SR results by iteratively predicting Gaussian noise during the denoising process, they are primarily designed for general images and struggle to effectively handle the unique characteristics and information present in LF images.
To address these limitations, we introduce LFSRDiff, the first diffusion-based LF image SR model, by incorporating the LF disentanglement mechanism. Our novel contribution includes the introduction of a disentangled U-Net for diffusion models, enabling more effective extraction and fusion of both spatial and angular information within LF images. Through comprehensive experimental evaluations and comparisons with the state-of-the-art LF image SR methods, the proposed approach consistently produces diverse and realistic SR results. It achieves the highest perceptual metric in terms of LPIPS. It also demonstrates the ability to effectively control the trade-off between perception and distortion. The code is available at \url{https://github.com/chaowentao/LFSRDiff}.

\end{abstract}

\section{Introduction}
\label{sec:intro}

Light field (LF) cameras can simultaneously record the spatial and angular information of the scene through a single snapshot, making them valuable for various applications such as depth estimation \cite{heber2016convolutional, shin2018epinet, tsai2020attention, chen2021attention, khan2021differentiable, wang2022occlusion, leistner2022towards, sheng2023lfnat}, view synthesis \cite{wu2018light, meng2019high, jin2020deep}, 3D reconstruction \cite{kim2013scene}, and virtual reality \cite{yu2017light}. 
However, due to the inherent trade-off between the spatial and angular LF cameras \cite{zhu2019revisiting}, the spatial resolution of LF images imposes a constraint on the accuracy of subsequent applications. In recent years, diverse LF image super-resolution (SR) methods \cite{yoon2017light, jin2020light, zhang2021end, zhang2019residual, wang2018lfnet, liang2022light, wang2022disentangling, wang2020spatial, yeung2018light, meng2019high, meng2020high, cheng2021light, cheng2022spatial, chen2022light, van2023light, wang2023ntire} have been developed to enhance the resolution of LF images, leading to significant advancements. Existing mainstream methods are mainly based on convolutional neural networks (CNNs) and adopt various mechanisms to extract and exploit the spatial and angular information of LF images, including adjacent-view combination \cite{yoon2017light}, view-stack integration \cite{jin2020light, zhang2021end, zhang2019residual}, bidirectional recurrent fusion \cite{wang2018lfnet}, spatial-angular disentanglement \cite{liang2022light, wang2022disentangling, wang2020spatial, yeung2018light, liang2023learning}, and 4D convolutions \cite{meng2019high, meng2020high}. These methods are typically supervised using $L_1$ or $L_2$ pixel-wise loss functions.

\begin{figure}[tb]
  \centering
   \includegraphics[width=\linewidth]{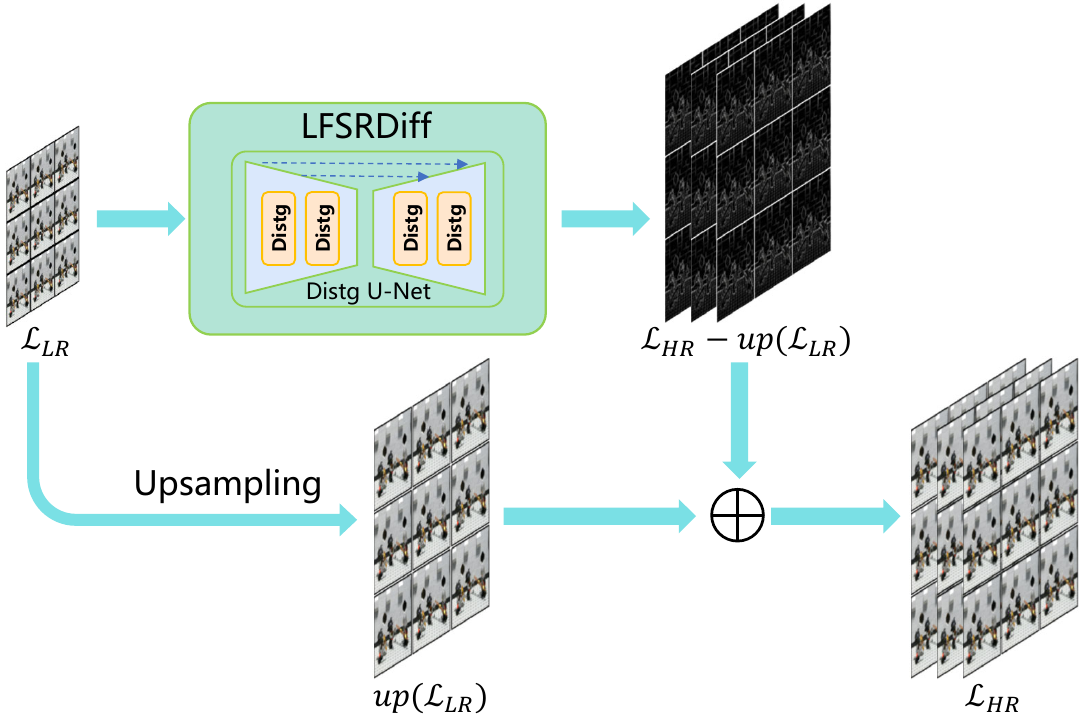}
   \caption{The diagram of LFSRDiff. Given an LR LF image, LFSRDiff is capable of learning the candidate SR distribution and generating stochastic multiple residuals, combining them with deterministic upsampled LR, to achieve diverse realistic SR results.}
   \label{fig: diagram}
\end{figure}

As widely acknowledged, LF image SR is an ill-posed problem, i.e., \textit{one-to-many} mapping, where numerous SR images can correspond to the same low-resolution (LR) image through downsampling. However, prevailing LF image SR methods tend to overlook the inherent nature of the LF image SR task whereby the use of PSNR-oriented loss gives rise to the issue of \textit{regression to the mean}. Consequently, these methods tend to produce deterministic outcomes, confining the output to a single SR result that is not only blurry but also lacks realism. 
Diffusion probabilistic models (Diffusion models) \cite{sohl2015deep} are built upon Markov chains that progressively introduce noise to the data. By learning the inverse process of iterative denoising, diffusion models can transform latent variables from simple distributions, such as Gaussian distributions, into data following complex distributions. Consequently, diffusion models are well-suited for addressing \textit{one-to-many} mapping problems and have demonstrated exceptional results in various tasks \cite{ho2020denoising, li2022srdiff, whang2022deblurring, rombach2022high, saharia2022image}. 

In this study, we revisit the nature of the LF image SR task with the objective of capturing the distribution of candidate SR images. This approach allows us to generate multiple diverse SR results for a given LR image, aligning more closely with the inherent nature of the SR task. 
We propose to introduce the diffusion model (LFSRDiff) to LF image SR, to the best of our knowledge, for the first time. However, the existing diffusion model component (U-Net) is primarily designed for general images and does not consider the characteristics of LF images where spatial and angular information are coupled together. Thus, the direct application of the diffusion model alone fails to deliver satisfactory results in LF image SR. Drawing inspiration from spatial-angular disentanglement work \cite{wang2022disentangling}, we incorporate the disentanglement mechanism into the diffusion model and propose disentangled U-Net (Distg U-Net) for learning noise. Direct learning of the SR distribution poses challenges in terms of model training difficulty, as well as the significant time and computational resources required during inference. To mitigate these issues, we adopt residual modeling \cite{he2016deep, li2022srdiff, whang2022deblurring} where the diffusion model learns the residuals between ground truth images and upsampled images. The approach can not only expedite model training but also yield better results. Fig.~\ref{fig: diagram} illustrates the diagram of LFSRDiff. The results of our method are more realistic and diverse than existing methods.

In summary, this paper makes the following contributions: (1) We revisit the nature of the LF image SR task and propose the first diffusion-based model for LF image SR. (2) We integrate the LF disentanglement mechanism into the diffusion model and propose a Distg U-Net for inverse process noise learning. (3) Compared with existing state-of-the-art LF image SR methods, our approach not only generates realistic and diverse results but also provides the flexibility to control the trade-off between perception and distortion.

\section{Related Work}
\label{sec: related}

\subsection{Light Field Image Super-Resolution}

In recent years, CNNs have been extensively applied in LF image SR and have demonstrated superior performance compared to traditional methods. The core in LF image SR networks is the effective aggregation of complementary information from adjacent views. 
LFCNN \cite{yoon2017light} is the first method to adopt CNN to learn the correspondence between stacked sub-aperture images (SAIs). SAIs are first super-resolved individually by SRCNN \cite{dong2014learning} and then fine-tuned in pairs to increase spatial and angular resolution. Jin et al. \cite{jin2020light} proposed an \textit{all-to-one} LF image SR method and performed structural consistency regularization to preserve the disparity structure. Apart from these two-stage methods \cite{yoon2017light, jin2020light}, which are time-consuming and computationally intensive, numerous one-stage networks have also been proposed for LF image SR. Yeung et al. \cite{yeung2018light} designed a spatial-angular separable (SAS) convolution to approximate 4D convolution and characterize the sub-pixel relationship of the LF 4D structure. Wang et al. \cite{wang2018lfnet} proposed a bidirectional recurrent network by extending BRCN \cite{huang2015bidirectional} into LF. Meng et al. \cite{meng2019high} proposed a densely connected network with 4D convolutions to explicitly learn the spatial angular correlation encoded in 4D LF data. Zhang et al. \cite{zhang2021end, zhang2019residual} grouped LF into four different branches according to specific angular directions and used four sub-networks to model multi-directional spatial angular correlation. Wang et al. \cite{wang2022disentangling} disentangled the spatial-angular correlation and proposed the method DistgSSR for LF image SR. Liang et al. \cite{liang2022light} applied the Transformer to LF image SR for the first time and designed a Spatial Transformer and Angular Transformer to incorporate spatial and angular information respectively. Recently, Wang et al. \cite{wang2020light} applied deformable convolution to LF images to solve the disparity problem in LF spatial SR. Liang et al. \cite{liang2023learning} proposed a Transformer-based method EPIT for learning the non-local spatial angular correlation on the two-dimensional epipolar plane image (EPI) plane of LF images. 

However, these methods are generally trained based on $L_1$ or $L_2$ pixel-wise loss and encounter the \textit{regression-to-mean} problem, which leads to deterministic behavior, where these methods can only produce a single blurry and unrealistic result, overlooking the intrinsic complexity of the LF image SR task. In contrast, our proposed diffusion-based LF image SR method focuses on learning the distribution of SR images, which is more consistent with the nature of the LF image SR task.

\subsection{Diffusion Models in Image Synthesis}

Recently, diffusion models \cite{sohl2015deep} are based on the Markov chains of denoising autoencoders, and have been shown to achieve impressive results in image synthesis \cite{ho2020denoising, song2020score, li2022srdiff, saharia2022image, whang2022deblurring, rombach2022high, dhariwal2021diffusion, ho2022cascaded}. Li et al. \cite{li2022srdiff} introduced the first diffusion-based single image super-resolution (SISR) model SRDiff, and have proven that it is feasible and promising to use the diffusion model to perform SISR tasks. Saharia et al. proposed SR3 \cite{saharia2022image}, which adopted the diffusion models to perform SISR tasks and produce competitive perception-based evaluation metrics. Whang et al. \cite{whang2022deblurring} proposed a framework for blind deblurring based on conditional diffusion models, trained a stochastic denoiser that refines the output of a deterministic predictor and can produce multiple plausible reconstructions for a given blurry input. Rombach et al. \cite{rombach2022high} proposed a latent diffusion model (LDM), which can significantly improve the training efficiency and sampling efficiency of denoising diffusion models without reducing model quality. The cascaded diffusion model \cite{ho2022cascaded} consisted of a pipeline of multiple diffusion models that generate images of increasing resolution, starting with a standard diffusion model at the lowest resolution, followed by one or more SR diffusion models to add higher resolution details.

However, the U-Net, as the core of the diffusion model, is originally designed for general images and does not fully consider the characteristics of LF images. To address this limitation, we integrate the LF image disentanglement mechanism with the U-Net network and propose Distg U-Net, which is used to extract the spatial and angular information of the LF image, respectively. Similar to \cite{li2022srdiff, whang2022deblurring}, we adopt residual modeling to expedite both model training and inference processes.


\begin{figure}[tb]
  \centering
   \includegraphics[width=\linewidth]{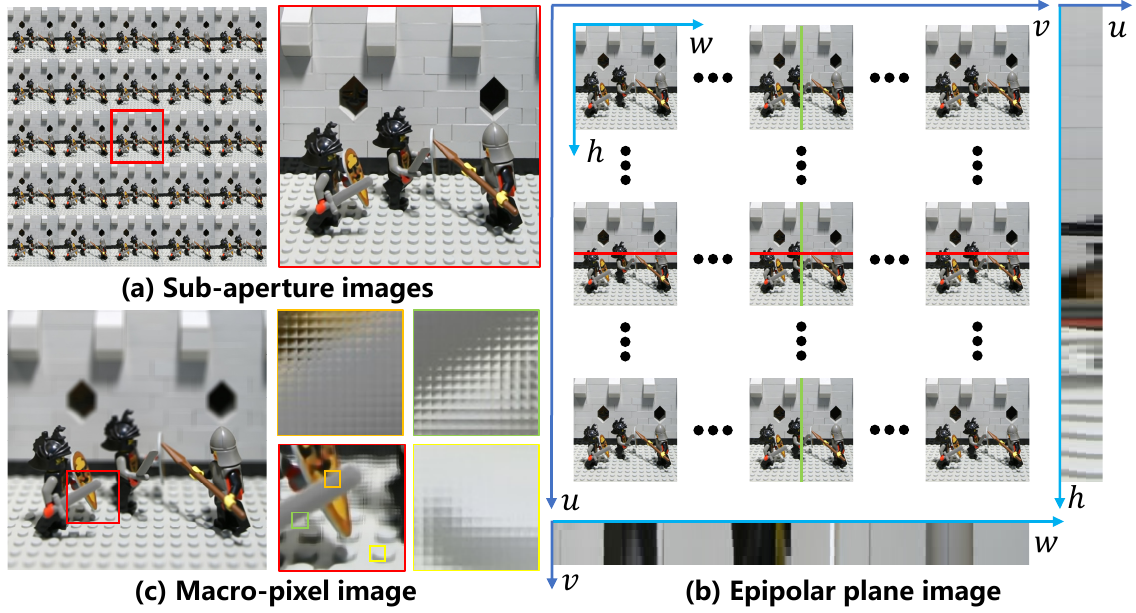}
   \caption{The representations of an LF scene \textit{Lego Knights} in 5$\times$5 views. }
   \label{fig: LF_vis}
\end{figure}

\section{LFSR Diffusion Model}
\label{sec:method}

\subsection{Preliminary}

We adopt the two-plane LF parameterization model \cite{levoy1996light} to represent an LF image, which can be formulated as a 4D function $\mathcal{L} (u,v,h,w)\in \mathbb{R}^{U\times V\times H\times W}$, where $U$ and $V$ stand for angular dimensions, where $H$ and $W$ stand for spatial dimensions. For better visualization, we can fix two dimensions in the 4D LF representation $\mathcal{L} (u,v,h,w)$. By fixing $u = u^{*}$ and $v = v^{*}$, we can get a certain sub-aperture image (SAI) $\mathcal{L} (u^{*},v^{*},h,w)$, as shown in Fig.~\ref{fig: LF_vis}(a). We can also fix $v = v^{*}$ and $w = w^{*}$ to obtain the horizontal EPI $\mathcal{L} (u,v^{*},h,w^{*})$, as shown in the horizontal direction in Fig.~\ref{fig: LF_vis}(b). Another form is the macro-pixel image (MacPI) that by fixing $h = h^{*}$ and $w = w^{*}$, a specific macro-pixel $\mathcal{L} (u,v,h^{*},w^{*})$ can be obtained, as shown in Fig.~\ref{fig: LF_vis}(c).


\begin{figure*}[tb]
  \centering
   \includegraphics[width=0.98\linewidth]{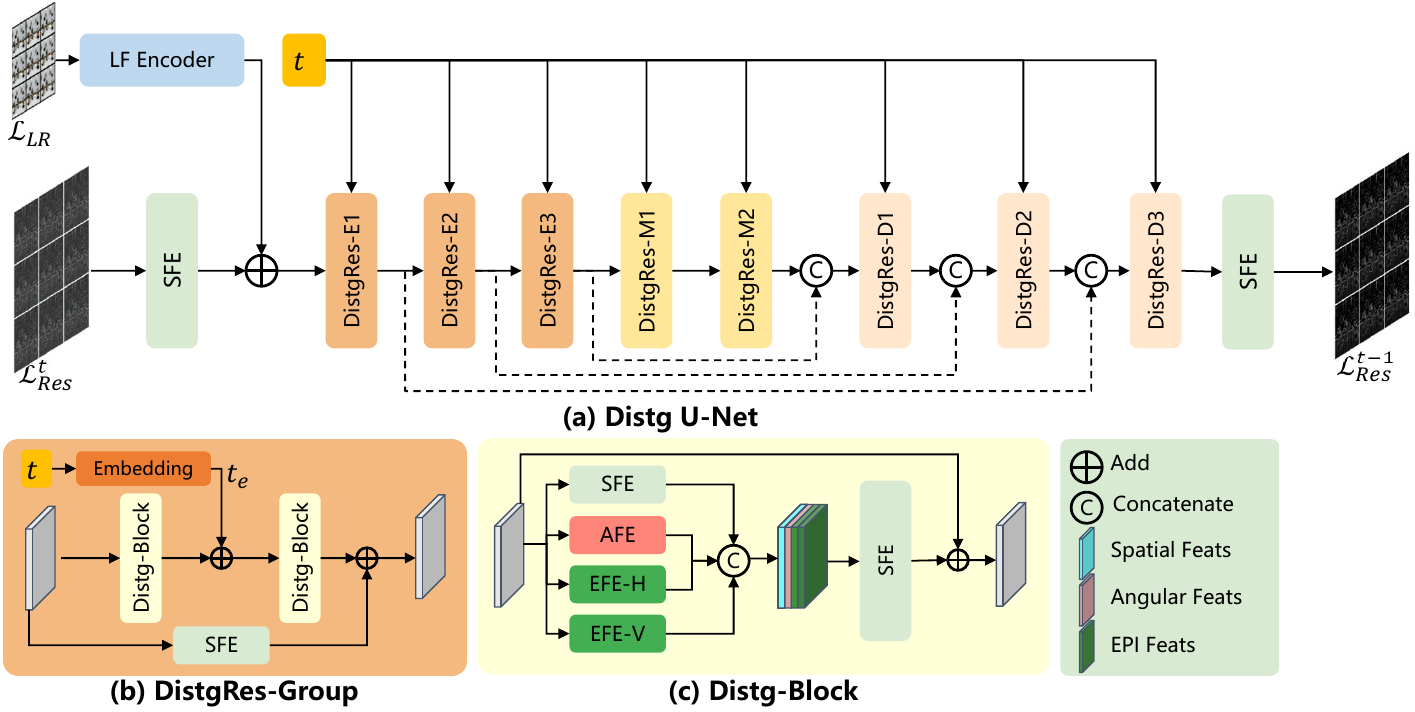}
   \caption{The architecture of Distg U-Net. Here, a 3$\times$3 LF is employed to illustrate.}
   \label{fig: LF_distg_unet}
\end{figure*}

\subsection{Conditional Diffusion Models}

Our method takes the LR LF $\mathcal{L}_{LR}\in \mathbb{R}^{U\times V\times H\times W}$ as inputs, and generate the SR LF $\mathcal{L}_{SR}\in \mathbb{R}^{U\times V\times \alpha H\times \alpha W}$ as outputs, where $\alpha$ stands for the scale factor of SR. This process can be described as a conditional distribution $p(\mathcal{L}_{SR} | \mathcal{L}_{LR})$, which is a \textit{one-to-many} mapping since many candidate SR LFs can correspond to the same LR LF through downsampling. Existing methods are supervised by $ L_1$ or $L_2$ pixel-wise loss, which is equivalent to regressing towards the mean of multiple candidate SR. This approach often yields blurred and unrealistic model outputs, neglecting the nature of the SR task. In contrast, we aim to directly learn the distribution $p(\mathcal{L}_{SR} | \mathcal{L}_{LR})$ through conditional diffusion models.


Following \cite{ho2020denoising, sohl2015deep}, a \textit{forward} diffusion process $q$ based on the Markov chain can be defined that gradually adds Gaussian noise to an LF HR image $\mathcal{L}_{HR}^{0}$ over $T$ iterations:

\begin{equation}
q(\mathcal{L}_{HR}^{t}|\mathcal{L}_{HR}^{t-1}) = \mathcal{N} (\mathcal{L}_{HR}^{t};\sqrt{\alpha_t } \mathcal{L}_{HR}^{t-1},(1-\alpha _t)\mathbf{I}   )
\label{eq: forward_1}
\end{equation}

\noindent where the $0<\alpha_t <1$ for all $(1,\dots, T)$ is the hyperparameter that controls the variance of Gaussian noise added at each iteration. Based on the Markov chain,  we can also represent the distribution of $\mathcal{L}_{HR}^{t}$ given $\mathcal{L}_{HR}^{0}$:

\begin{equation}
q(\mathcal{L}_{HR}^{t}|\mathcal{L}_{HR}^{0}) = \mathcal{N} (\mathcal{L}_{HR}^{t};\sqrt{\overline \alpha_t } \mathcal{L}_{HR}^{0},(1-\overline \alpha_t)\mathbf{I}  )
\label{eq: forward_2}
\end{equation}

\noindent where $\overline \alpha_t=\prod_{j=1}^{t} \alpha_j$. Furthermore, the posterior distribution of $q(\mathcal{L}_{HR}^{t-1} | \mathcal{L}_{HR}^{t}, \mathcal{L}_{HR}^{0})$ can be derived as:

\begin{equation}
q(\mathcal{L}_{HR}^{t-1}|\mathcal{L}_{HR}^{t},\mathcal{L}_{HR}^{0})  = \mathcal{N} (\mathcal{L}_{HR}^{t-1};\mu_t,\sigma^2_t\mathbf{I})
\label{eq: forward_post}
\end{equation}

\noindent where $\mu_t = \frac{\sqrt{\overline \alpha_{t-1} }(1-\alpha _t)}{1-\overline\alpha _t } \mathcal{L}_{HR}^{0}+\frac{\sqrt{ \alpha_t }(1-\overline \alpha_{t-1} )}{1-\overline\alpha _t } \mathcal{L}_{HR}^{t}$ and $\sigma^2_t = \frac{(1-\overline \alpha_{t-1})(1-\alpha _t)}{1-\overline\alpha _t }$.

The \textit{reverse} diffusion step in Eq.~\eqref{eq: forward_post} allows reversing a single diffusion step by sampling a slightly less noisy image $\mathcal{L}_{HR}^{t-1}$ from $\mathcal{L}_{HR}^{t}$. Note that the forward diffusion process is fixed and iteratively adds $T$ times Gaussian noise to the HR LF image $\mathcal{L}_{HR}^{0}$ until the noisy LF image $\mathcal{L}_{HR}^{t}$is indistinguishable from Gaussian noise. However, it is impossible for \textit{reverse} diffusion step to access $\mathcal{L}_{HR}^{0}$, which is what we aim to generate. So we can define a denoising network (generally U-Net) $f_\theta$ to learn the estimated distribution $p(\mathcal{L}_{HR}^{t-1} | \mathcal{L}_{HR}^{t})$ approximate the posterior distribution of $q(\mathcal{L}_{HR}^{t-1} | \mathcal{L}_{HR}^{t}, \mathcal{L}_{HR}^{0})$. In practice, we follow the \cite{ho2020denoising} to predict Gaussian noise $\epsilon$ by  $\mathcal{L}_{HR}^{t}=\sqrt{\overline{\alpha}_t}\mathcal{L}_{HR}^{0}+\sqrt{(1-\overline{\alpha}_t)}\epsilon$ where $\epsilon \sim \mathcal{N}(0,\mathbf{I}) $. 

For \textit{conditional} diffusion models, we need to introduce the condition $\mathcal{L}_{LR}$ into the denoising network $f_\theta$ to control model outputs. Similarly to \cite{whang2022deblurring, saharia2022image, li2022srdiff}, the training objective $L_{direct}(\theta )$  can be written as:

\begin{equation}
\left \| \epsilon - f_\theta (\sqrt {\overline \alpha _t}\mathcal{L}_{HR}^{0}+\sqrt{(1-\overline{\alpha}_t)}\epsilon,t,\mathcal{L}_{LR})\right \|_1
\label{eq: rev_obj_dir}
\end{equation}

\noindent \textbf{Residual Modeling.} 
Training a diffusion model to obtain the final HR LF image $\mathcal{L}_{HR}^{0}$ directly from Gaussian noise is difficult and inference is very time-consuming since needs a large $T$ iteration. Inspired by residual modeling \cite{he2016deep, li2022srdiff, whang2022deblurring}, we change the objective of the diffusion model to predict the residual between HR LF image $\mathcal{L}_{HR}^{0}$ and upsampled LR LF image $up(\mathcal{L}_{LR})$. The residual objective $L_{res}(\theta )$ can be written as:

\begin{equation}
 \left \| \epsilon - f_\theta (\sqrt {\overline \alpha _t}(\mathcal{L}_{HR}^{0}-up(\mathcal{L}_{LR}))+\sqrt{(1-\overline{\alpha}_t)}\epsilon,t,\mathcal{L}_{LR})\right \|_1
\label{eq: rev_obj_res}
\end{equation}


\begin{figure}[tb]
  \centering
   \includegraphics[width=\linewidth]{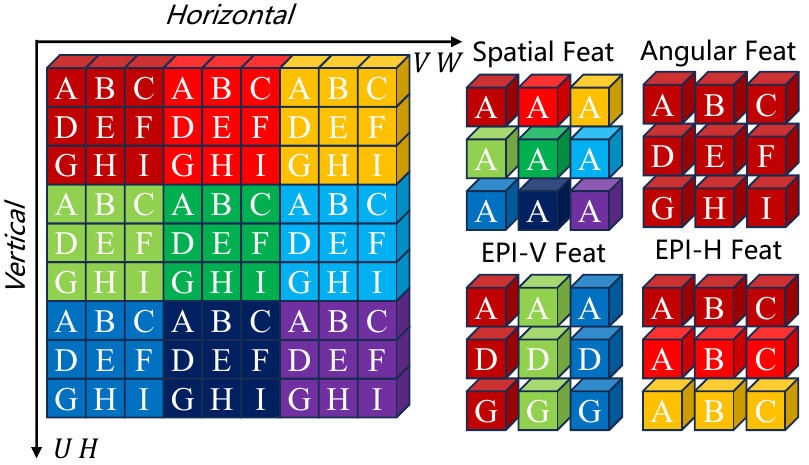}
   \caption{The exhibition of spatial, angular, and EPI features based on a toy LF MacPI representation with $U=V=3$ and $H=W=3$. Here, the same color represents the same angular domain information, and the same letter represents the same spatial domain information. }
   \label{fig: LF_distg}
\end{figure}

\subsection{Distg U-Net}

Since LF images contain both spatial information and angle information that are entangled with each other, the original U-Nets in the existing diffusion models \cite{li2022srdiff, whang2022deblurring, saharia2022image} are designed for general images and cannot adeptly handle LF images. To address this limitation, we incorporate the LF disentanglement mechanism \cite{wang2022disentangling} into the network design of the diffusion model. First, we introduce the LF disentanglement mechanism.

\subsubsection{Disentanglement Mechanism in LF}
\label{sec: distg}

As shown in Fig.~\ref{fig: LF_distg}, the LF disentanglement mechanism \cite{wang2022disentangling} obtains spatial, angular, and EPI features according to different combinations of LF image pixels. Furthermore, by setting different convolution kernel sizes, strides, and dilated rates, the corresponding features of the LF image can be extracted. Here, we employ three types of feature extractors for feature extraction, $U=V=A$.

\textit{Spatial Feature Extractor (SFE)} is a convolution with a kernel size of 3$\times$3, a stride of 1, and a dilation of $A$.

\textit{Angular Feature Extractor (AFE)} is a convolution with a kernel size of 3$\times$3, a stride of $A$, and a dilation of 1. 

\textit{EPI Feature Extractors (EFE)} need to extract horizontal and vertical EPI features. For horizontal EPI (EPI-H) features, we design EFE-H as a convolution with a kernel size of 1$\times A^2$, a vertical stride of 1, and a horizontal stride of $A$. EFE-V adopts a similar approach.

\subsubsection{Network Design}

An overview of our Distg U-Net is shown in Fig.~\ref{fig: LF_distg_unet}(a). According to Eq.~\eqref{eq: rev_obj_res}, the input to Distg U-net $f_\theta$ is timestep $t$, LR LF image $\mathcal{L}_{LR}$ and the noisy residual HR LF image $\mathcal{L}_{Res}^{t}$ and generates the residual LF image $\mathcal{L}_{Res}^{t-1}$ by predicting the noise at timestep $t$. 

Firstly, we use the SFE designed in Sec.~\ref{sec: distg} to extract $\mathcal{L}_{Res}^{t}$ spatial feature. Then, the spatial feature is added with LR LF $\mathcal{L}_{LR}$ feature, extracted by a generalized LF Encoder. Follow \cite{ho2020denoising, li2022srdiff}, we convert timestep $t$ into a timestep embedding $t_e$ using sinusoidal Positional Encoding commonly proposed in Transformer \cite{vaswani2017attention}. The main components of Distg U-Net contain three DitsgRes-Groups in the encoder (DistgRes-E), two DitsgRes-Groups in the bottleneck (DistgRes-M), and three DitsgRes-Groups in the decoder (DistgRes-D). Skip connections are used between the encoder and the decoder at the same level. Each DitsgRes-Group (see Fig.\ref{fig: LF_distg_unet}(b)) consists of a residual spatial convolution, two disentangling blocks (Distg-Blocks). The timestep embedding $t_e$ is added to the extracted feature of the first Distg-Blocks through spatial replication. DistgRes-E and DistgRes-D have an extra downsampling or upsampling operation.


\noindent \textbf{Distg-Block.} The Distg-Block is the basic module of our Distg U-Net based on the LF disentanglement mechanism mentioned in Sec.~\ref{sec: distg} (as shown in Fig.\ref{fig: LF_distg_unet}(c)). Specifically, we employ SFE, AFE, EFE-V, and EFE-H to disentangle spatial, angular, and EPI features, respectively. Then, angular features and EPI features are upsampled and concatenated with spatial features. Finally, an SFE is adopted to fuse concatenated features and output fused features. Note that the Distg-Block in the LF encoder contains a residual connection, while DitsgRes-Group does not use a residual connection since the feature map is downsampled.

\noindent \textbf{LF Encoder.} The LR LF feature is encoded by the LF Encoder and is added to each reverse step to guide the generation to the corresponding HR LF image. In this paper, we choose the EPIT \cite{liang2023learning} as the LF Encoder, which is robust to disparity variations. Our LFSRDiff is generalized to different LF Encoders and we also incorporate Distg U-Net with other different LF Encoders (see Sec.~\ref{sec: generalization}).


\section{Experiments}
\label{sec:experiments}

In this section, we first introduce the dataset, implementation details, and evaluation metrics. Then, we quantitatively and qualitatively compare our approach with state-of-the-art methods. Next, we compare the performance of different LF image SR methods in real LF scenes. Finally, we verify the effectiveness and robustness of our method through a series of ablation experiments. 

\subsection{Experimental Settings}

\noindent \textbf{Datasets.} Following previous methods \cite{wang2020light, wang2020spatial, liu2021intra, wang2022detail, liang2022light}, we use five mainstream LF image datasets, i.e., EPFL \cite{rerabek2016new}, HCINew \cite{honauer2016dataset}, HCIold \cite{wanner2013datasets}, INRIA \cite{le2018light}, and STFgantry \cite{vaish2008new} for our LFSR experiments and convert the input LF image to YCbCr color space and only super-resolve the Y channel of the image, upsampling the Cb and Cr channel images bicubic. We extract the center 5$\times$5 with angular resolution 9$\times$9 LF images in these datasets for training and testing. 
For the convenience of training, we crop the LF image into LF patches with a size of 128$\times$128 and a stride of 64 and perform bicubic downsampling 4$\times$ to obtain LR patches for 4$\times$SR experiments.

\noindent \textbf{Implementation Detail.} We adopt the same diffusion settings for fair comparisons with SRDiff \cite{li2022srdiff}: The timestep $T$ is set to 100 for the diffusion process and the reverse process. The noise schedule follows \cite{nichol2021improved, li2022srdiff}  and uses a cosine noise schedule. We adopt a two-stage training strategy. First, we pre-train the LF Encoder using an $L_1$ loss for the sake of efficiency. Then, we fix the LF Encoder and use the training loss mentioned in Eq.\eqref{eq: rev_obj_res} for training Distg U-Net.
We adopt Adam \cite{kingma2014adam} as the optimizer, set the batchsize to 4, the learning rate to $2 \times 10^{-4}$ and decrease it by half every 100k iterations. 
We employ random horizontal flipping, vertical flipping, and 90-degree rotation for data augmentation. Our Distg U-Net is implemented in the PyTorch framework and trained with 200k iterations using an Nvidia A100 GPU. The procedure of training and inference of our LFSRDiff can be referred to in the supplementary material.


\noindent \textbf{Evaluation Metrics.} We adopt the well-known distortion-based metrics (PSNR and SSIM \cite{wang2004image}). Since these metrics are not consistent with human perception, we also introduce the popular LPIPS \cite{zhang2018unreasonable} as perceptual evaluation. Following previous methods \cite{wang2020light, wang2020spatial, liu2021intra, wang2022detail, liang2022light}, when calculating the average metric for a dataset, we first average the metric scores for all SAIs for each scene separately and then average the metric scores for all scenes of the dataset.

\noindent \textbf{Sample Averaging.} Since our model is trained to learn the posterior distribution $p(\mathcal{L}_{SR} | \mathcal{L}_{LR})$, we can produce multiple results and support geometric self-ensemble \cite{whang2022deblurring, lim2017enhanced}. The sampling averaging strategy achieves a trade-off between distortion and perception and is equivalent to \textit{regression to the mean}, which is consistent with the learning objectives of existing SR methods \cite{wang2020light, wang2020spatial, liu2021intra, wang2022detail, liang2022light}. To make a better comparison with other methods in terms of distortion metrics, we also evaluate the sampling averaging (SA) metrics.

\begin{table*}[ht]
\centering
\caption{Quantitative comparison of different SR methods in terms of distortion (PSNR) and perception (LPIPS) metrics. The \colorbox {blue!20}{best values} and \colorbox {green!20}{second-best values} for each metric are color-coded. Our method achieves a trade-off between distortion and perception.}
\renewcommand\arraystretch{1.0}
\resizebox{2 \columnwidth}{!}{
\begin{tabular}{lcccccccccccc}
\toprule
\multicolumn{1}{c}{\textbf{Method}}   &  \multicolumn{2}{c}{\textbf{EPFL}} & \multicolumn{2}{c}{\textbf{HCInew}} & \multicolumn{2}{c}{\textbf{HCIold}} & \multicolumn{2}{c}{\textbf{INRIA}} & \multicolumn{2}{c}{\textbf{STFgantry}} & \multicolumn{2}{c}{\textbf{Average}} \\ 
\cline{2-13} 
\multicolumn{1}{c}{4$\times$}  & PSNR$\uparrow$ & LPIPS$\downarrow$ & PSNR$\uparrow$ & LPIPS$\downarrow$ & PSNR$\uparrow$ & LPIPS$\downarrow$ & PSNR$\uparrow$ & LPIPS$\downarrow$ & PSNR$\uparrow$ & LPIPS$\downarrow$ & PSNR$\uparrow$ & LPIPS$\downarrow$  \\
\midrule
 Bicubic & 25.14 & 0.4617 & 27.61 & 0.4983 & 32.42 & 0.3583 & 26.82 & 0.4368 & 25.93 & 0.4633 & 27.58 & 0.4437 \\
 VDSR \cite{kim2016accurate} (2016) & 27.25 & 0.2795 & 29.31 & 0.3191 & 34.81 & 0.1990 & 29.19 & 0.2611 & 28.51 & 0.1915 &  29.81 & 0.2501 \\
 EDSR \cite{lim2017enhanced} (2017) & 27.84 & 0.2606 & 29.60 & 0.3117 & 35.18 & 0.1977 & 29.66 & 0.2497 & 28.70 & 0.1671 &  30.19 & 0.2374 \\
 RCAN \cite{zhang2018image} (2018)  & 27.88 & 0.2594 & 29.63 & 0.3088 & 35.20 & 0.2007 & 29.76 & 0.2471 & 28.90 & 0.1548 & 30.36 & 0.2342 \\
 SRDiff\cite{li2022srdiff} (2022)  & 26.94 & 0.2174 & 28.73 & 0.2702 & 34.16 & 0.1590 & 28.93 & 0.2030 & 27.45 & 0.1454 & 29.24 & 0.1990  \\
 \midrule
 resLF \cite{zhang2019residual} (2019) & 28.27 & 0.2351 & 30.73 & 0.2556 & 36.71 & 0.1380 & 30.34 & 0.2244 & 30.19 & 0.1348 & 31.24 & 0.1976 \\
 LFSSR \cite{yeung2018light} (2018)  & 28.27 & 0.2123 & 30.72 & 0.2458 & 36.70 & 0.1331 & 30.31 & 0.2078 & 30.15 & 0.1235 & 31.52 & 0.1845 \\
 LF-ATO \cite{jin2020light} (2020) & 28.52 & 0.2117 & 30.88 & 0.2534 & 37.00 & 0.1343 & 30.71 & 0.2035 & 30.61 & 0.1212 & 31.54 & 0.1848 \\
 LF-InterNet \cite{wang2020spatial} (2020)  & 28.67 & 0.2106 & 30.98 & 0.2444 & 37.11 & 0.1200 & 30.64 & 0.2083 & 30.53 & 0.1341 & 31.61 & 0.1835 \\
 LF-DFnet \cite{wang2020light} (2020) & 28.77 & 0.1948 & 31.23 & 0.2356 & 37.32 & 0.1213 & 30.83 & 0.1922 & 31.15 & 0.1031  & 31.86 & 0.1694 \\
 MEG-Net \cite{zhang2021end} (2021) & 28.74 & 0.2076 & 31.10 & 0.2307 & 37.28 & 0.1197 & 30.66 & 0.2021 & 30.77 & 0.1162  & 31.72 & 0.1753 \\
 LF-IINet \cite{liu2021intra} (2021) & 29.11 & 0.2020 & 31.36 & 0.2274 & 37.62 & 0.1126 & 31.08 & 0.1943 & 31.21 & 0.0978 & 32.06 & 0.1668 \\
 DPT \cite{wang2022detail} (2022) & 28.93 & 0.2073 & 31.19 & 0.2389 & 37.39 & 0.1216 & 30.96 & 0.1983 & 31.14 & 0.1032 & 31.93 & 0.1739 \\
 LFT \cite{liang2022light} (2022) & 29.25 & \cellcolor{green!20}0.1926 & \cellcolor{green!20}31.46 & 0.2260 & \cellcolor{green!20}37.63 & 0.1074 & 31.20 & 0.1889 & \cellcolor{green!20}31.86 & 0.0928 & 32.28 & 0.1615 \\
 DistgSSR \cite{wang2022disentangling} (2022) & 28.99 & 0.1965 & 31.38 & \cellcolor{green!20}0.2203 & 37.56 & \cellcolor{green!20}0.1061 & 30.99 & \cellcolor{green!20}0.1883 & 31.65 & \cellcolor{green!20}0.0843 & 32.12 & \cellcolor{green!20}0.1591 \\
 EPIT \cite{liang2023learning} (2023) & \textbf 29.34 & 0.1999 & \cellcolor{blue!20}31.51 & 0.2206 & \cellcolor{blue!20}37.68 & 0.1140 & 31.27 & 0.1893 & \cellcolor{blue!20}32.18 & \cellcolor{blue!20}0.0842 & \cellcolor{green!20}32.41 & 0.1616  \\
 \midrule
 LFSRDiff (Ours) & \cellcolor{green!20}29.38 & \cellcolor{blue!20} 0.1645 & 31.10 & \cellcolor{blue!20} 0.1819 & 37.13 & \cellcolor{blue!20} 0.1031 & \cellcolor{green!20}31.58 & \cellcolor{blue!20} 0.1535 & 31.58 & 0.0929 & 32.15 & \cellcolor{blue!20} 0.1392 \\
 LFSRDiff (Ours-SA) & \cellcolor{blue!20}29.62 & 0.2141 & 31.32 & 0.2366 & 37.45 & 0.1256 & \cellcolor{blue!20}31.86 & 0.2027 & 31.85 & 0.1077 & \cellcolor{blue!20}32.42 & 0.1773 \\
\bottomrule
\end{tabular}
}

\label{table: quantitative}
\end{table*}

\begin{figure*}[tb]
  \centering
   \includegraphics[width=\linewidth]{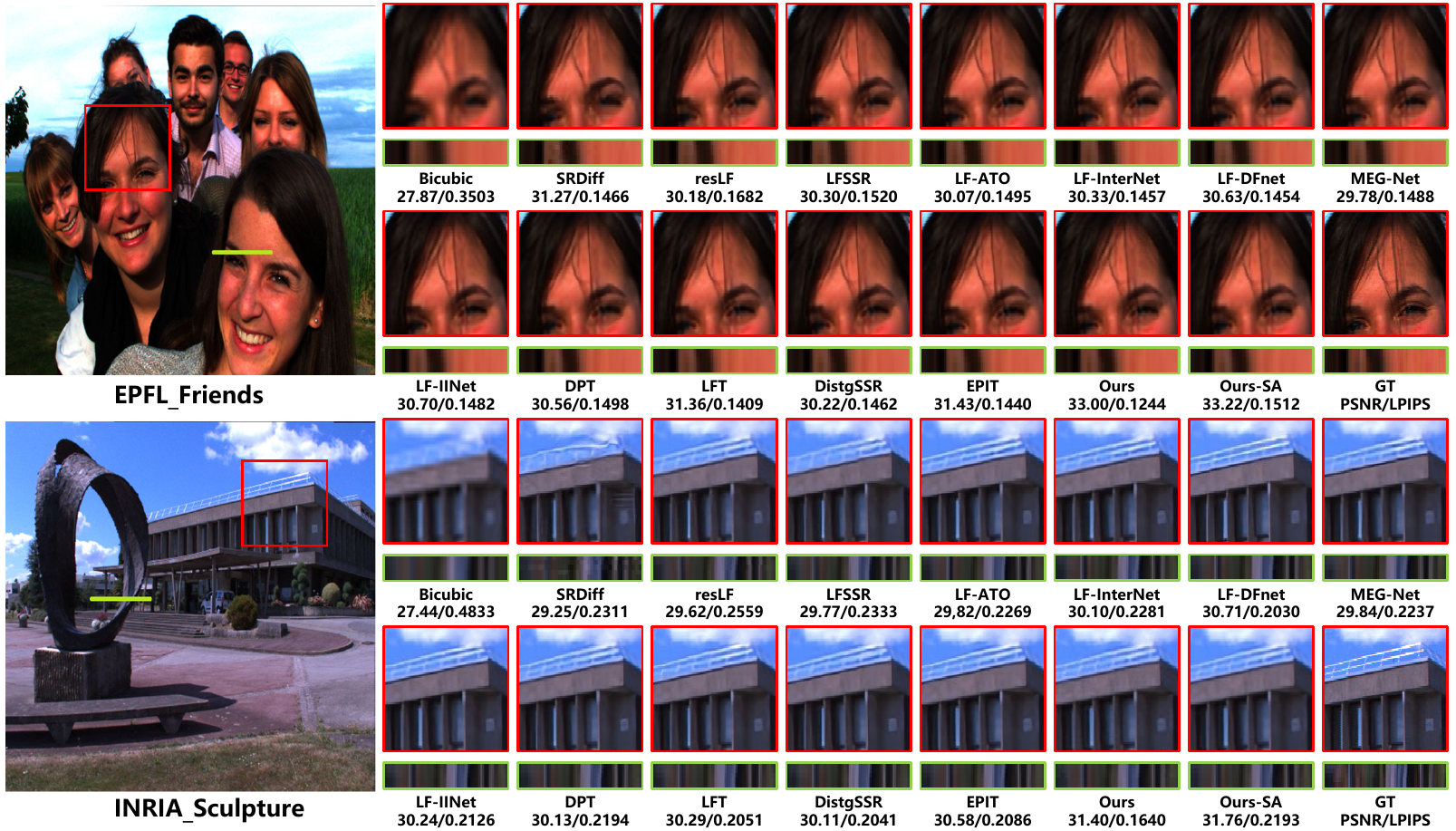}
   \caption{Qualitative comparison of different SR methods for 4$\times$SR. The super-resolved center view images and horizontal EPIs are shown. Our method adopts an iterative noise learning approach instead of pixel-wise loss minimization, which effectively avoids generating blurry output and achieves superior perceptual reconstruction of detailed textures. Best viewed zoom-in electronically. }
   \label{fig: qualitative}
\end{figure*}

\begin{figure}[tb]
  \centering
   \includegraphics[width=\linewidth]{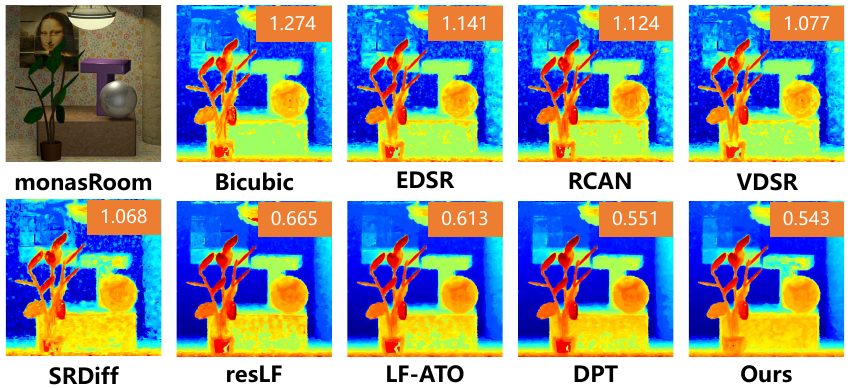}
   \caption{Depth estimation results of scene \textit{monasRoom} from HCIold \cite{wanner2013datasets} achieved by IGF \cite{sheng2017geometric} using 4$\times$ SR LF images produced by different SR methods. The MSE $\times$100 ($\downarrow$) is used as a quantitative evaluation metric and is labeled in the upper right corner.}
   \label{fig: compare_depth}
\end{figure}

\begin{figure}[tb]
  \centering
   \includegraphics[width=\linewidth]{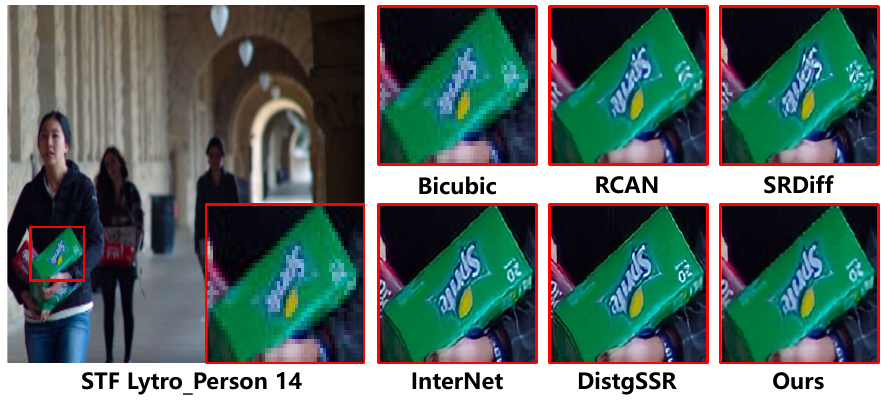}
   \caption{Visual comparison on a real-world scene \textit{Person 14} from STF Lytro \cite{SLFA}. Compared to other methods, our approach yields sharper edges, such as the letters in the enlargement region, and has improved overall visual perception. }
   \label{fig: compare_real}
\end{figure}

\subsection{Comparisons with state-of-the-art Methods}

We compare our method to 15 state-of-the-art SR methods, including 4 single image SR methods \cite{kim2016accurate, lim2017enhanced, zhang2018image, li2022srdiff} and 11 LF image SR methods \cite{zhang2019residual, yeung2018light, jin2020light, wang2020spatial, wang2020light, zhang2021end, liu2021intra, wang2022detail, liang2022light, wang2022disentangling, liang2023learning}. 

\noindent \textbf{Quantitative Results.} Table~\ref{table: quantitative} shows a quantitative comparison among LF image SR methods. 
Our LFSRDiff achieves the best perception score (i.e., LPIPS of 0.1392), nearly a 13\% reduction compared to DistgSSR \cite{wang2022disentangling}, and maintains a competitive distortion score (i.e., PSNR of 32.15) on five datasets for 4$\times$ SR tasks.
Moreover, our method through sampling average (Ours-SA) achieves the state-of-the-art distortion score (i.e., PSNR of 32.42). In other words, our method allows the flexibility to use a single model to control the trade-off between perception and distortion (see Sec~\ref{sec: trade-off}). Please refer to our supplemental material for additional quantitative comparisons.


\noindent \textbf{Qualitative Results.} Figure.~\ref{fig: qualitative} shows the qualitative results achieved by different methods for 4$\times$ SR. As can be seen from the zoomed-in areas, the SISR method (i.e., SRDiff) cannot reliably recover the missing details, such as the handrail area in the scene \textit{Sculpture}. Other LFSR methods tend to obtain higher PSNR, supervised by the $L_1$ loss, and lead to blurry results, such as hair texture and eyeballs in the scene \textit{Firends}. In contrast, our LFSRDiff can learn the SR LF image distribution by iteratively learning the noise and achieving the best perceptual quality. More qualitative results can be found in supplementary material.

\noindent \textbf{Angular Consistency.} As shown in Fig.~\ref{fig: qualitative}, compared with other SR methods, the horizontal EPI of our method can maintain sharper and clearer lines with fewer artifacts. The slope of these lines is related to the depth values, which demonstrates that our method can preserve the LF disparity structure well. Moreover, we also evaluate the angular consistency through a depth estimation algorithm IGF \cite{sheng2017geometric}. As shown in Fig.~\ref{fig: compare_depth}, the results of depth estimation by SISR methods contain much noise and have higher MSE $\times$100 error, which indicates SISR methods do not consider the angular information in LF image.
The utilization of LF images generated by our method yields a substantial improvement in the accuracy of depth estimation. The resulting depth map exhibits visually clear quality and demonstrates the lowest Mean Squared Error (MSE) $\times$100 error compared to alternative methods.
This further underscores the superior angular consistency achieved by our method. 

\noindent \textbf{Performance on Real Scenes.} We compare different methods on real scenes from STF Lytro \cite{SLFA}. As shown in Fig.~\ref{fig: compare_real}, some SR methods, such as RCAN \cite{zhang2018image} and SRDiff \cite{li2022srdiff}, result in blurry and lack of realism. The results obtained from InterNet \cite{wang2020light} and DistgSSR \cite{wang2022disentangling} methods also appear noticeably unrealistic.
In contrast, our method produces results with clearer edges and enhanced visual perception compared to other methods. This observation affirms that our approach achieves the expected goals with superior robustness. 

\subsection{Ablation Study}
In this subsection, we first compare the performance of different variants of Distg U-Net, then verify the generalization ability of Distg U-Net by using different LF Encoder networks. Next, we demonstrate the advantages of residual modeling. Finally, we show the sample diversity and sample averaging to achieve the trade-off between perception and distortion.

\begin{table}[tb]
\centering
\caption{PSNR, SSIM, and parameters achieved by different Distg U-Net (DU-Net) Variants for 4$\times$SR.}
\renewcommand\arraystretch{1.0}
\resizebox{\columnwidth}{!}{
\begin{tabular}{l|ccc|cc|c|cc}
\toprule
 & \textbf{Spa.} & \textbf{Ang.} & \textbf{EPI} & \textbf{ch.} & \textbf{dim. mults.} & \textbf{Params.} & \textbf{PSNR} & \textbf{SSIM} \\
 \midrule
U Net & - & - & - & 32 & 1\textbar2\textbar3\textbar4 & 2.91M & 31.07 & 0.9281 \\
\midrule
 
\multirow{6}{*}{DU-Net} & \checkmark &  &  & 32 & 1\textbar1\textbar2\textbar2 & 2.55M & 31.21 & 0.9325 \\
 &  & \checkmark &  & 32 & 1\textbar1\textbar2\textbar2 & 2.56M & 30.76 & 0.9250 \\
 & \checkmark & \checkmark &  & 32 & 1\textbar1\textbar2\textbar2 & 2.57M & 31.32 & 0.9332 \\
 & \checkmark & \checkmark & \checkmark & 32 & 1\textbar1\textbar1\textbar1 & 2.13M & 31.52 & 0.9332 \\
 & \checkmark & \checkmark & \checkmark & 32 & 1\textbar1\textbar2\textbar2 & 5.69M & 31.84 & 0.9398 \\
 & \checkmark & \checkmark & \checkmark & 64 & 1\textbar1\textbar1\textbar1 & 8.50M & 32.15 & 0.9386 \\

 \bottomrule
 
\end{tabular}
}
\label{table: variats}
\end{table}

\begin{table}[tb]
\centering
\caption{PSNR results achieved by several combinations between LF Encoders and U-Net or Distg U-Net (DU-Net) for 4$\times$SR.}
\renewcommand\arraystretch{1.0}
\resizebox{\columnwidth}{!}{
\begin{tabular}{l|ccccc|c}
\toprule
\textbf{Combinations} & \textbf{EPFL} & \textbf{HCInew} & \textbf{HCIold} & \textbf{INRIA} & \textbf{STFgtr} & \textbf{Avg.} \\
\midrule
InterNet+U-Net    & 28.15 & 29.78 & 36.16 & 30.49 & 29.26 & 30.77 \\
InterNet+DU-Net   & 28.94 & 30.54 & 36.41 & 31.11 & 29.83 & 31.37 \\ 
\midrule
DistgSSR+U-Net    & 28.17 & 30.12 & 36.23 & 30.58 & 29.90 & 31.00 \\
DistgSSR+DU-Net   & 28.86 & 30.61 & 36.42 & 31.14 & 30.36 & 31.48  \\ 
\midrule
EPIT+U-Net   & 28.19 & 30.09 & 36.23 & 30.60 & 30.22 & 31.07 \\
EPIT+DU-Net  & 28.62 & 30.78 & 36.60 & 30.64 & 30.91 & 31.52 \\ 
\bottomrule
\end{tabular}
}
\label{table: general}
\end{table}

\subsubsection{Distg U-Net Variants} We conduct experiments with different variants of Distg U-Net and compare them with the original U-Net \cite{li2022srdiff}. We control the parameters of different variants to be approximately equal to ensure a fair comparison. As shown in Table~\ref{table: variats}, using the disentanglement mechanism to extract spatial features alone, the metrics are already better than the original U-Net network. Furthermore, by exploring different combinations of spatial features, angular features, and EPI features respectively, we find that the combination of all three features achieves the best results. The feature channel dimension is further increased, and the results will be improved accordingly.

\subsubsection{Distg U-Net Generalization}
\label{sec: generalization}
We compare Distg U-Net with original U-Net \cite{li2022srdiff} in different LF Encoder networks, including InterNet \cite{wang2020spatial}, DistgSSR \cite{wang2022disentangling} and EPIT \cite{liang2023learning}. 
Specifically, we remove the last upsampling convolution layer of these models and use the extracted features as conditional information.
As shown in Table~\ref{table: general}, the Distg U-Net outperforms the original U-Net in all LF Encoders by a large margin (about average 0.5 dB PSNR). It demonstrates the generalization of our Distg U-Net.

\begin{figure}[tb]
  \centering
   \includegraphics[width=\linewidth]{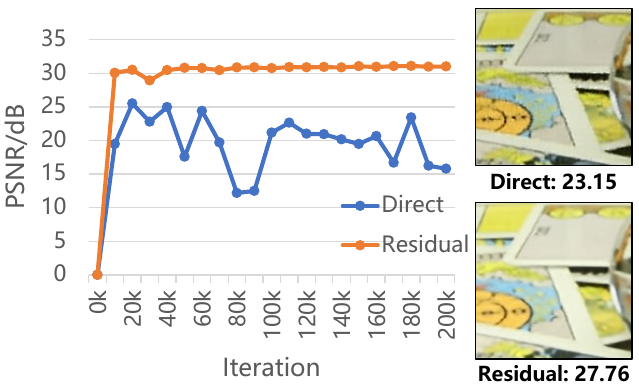}
   \caption{Training curves for direct learning and residual modeling. The right column compares the crop on a \textit{Cards} from the STFgantry \cite{vaish2008new}. Residual modeling has a more stable training curve and achieves higher PSNR than direct learning, resulting in better visual quality. }
   \label{fig: residual}
\end{figure}

\subsubsection{Residual Modeling}

We explore the performance of residual modeling and direct learning. As shown in Fig.\ref{fig: residual}, direct learning exhibits an unstable training process and produces outputs that are not realistic and contain much noise. In contrast, residual modeling achieves higher PSNR (i.e., 27.76 vs. 23.15) and better visual perception with a more stable training process.

\begin{figure}[tb]
  \centering
   \includegraphics[width=\linewidth]{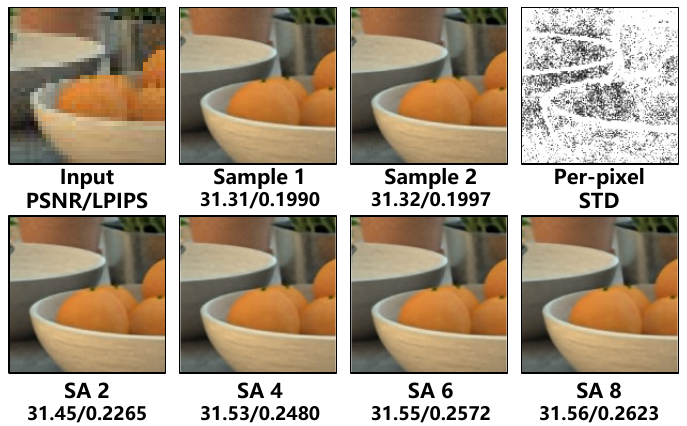}
   \caption{Illustration of diverse samples, per-pixel standard deviation (STD), and sample averaging (SA) strategy for the crop on a  scene \textit{Herbs} from HCINew \cite{honauer2016dataset}.}
   \label{fig: diversity}
\end{figure}

\subsubsection{Trade-off between Perception and Distortion}
\label{sec: trade-off}

We also investigate the impact of the sample averaging strategy on the performance. 
As depicted in Fig.~\ref{fig: diversity}, the absence of sample averaging leads to significantly higher per-pixel standard deviations (i.e., diversity) and maintaining perceptual quality, aligning with our objective of learning SR distribution.
On the other hand, with more samples used in sample averaging, the ensemble result tends to be closer to the mean and has a better distortion metric but worse perceptual quality. Therefore, the sample averaging strategy can be adjusted to achieve a favorable trade-off between perception and distortion, depending on specific requirements.




\section{Conclusion and Future Work}
\label{sec: conclusion} 

In this paper, we have proposed a novel diffusion-based network for LF image SR. By integrating the LF disentanglement mechanism with the diffusion model, we propose Distg U-Net for inverse process noise learning. Extensive experimental results consistently demonstrated the superior capability of our method in generating more realistic and diverse results than existing LF image SR methods.

While our proposed approach demonstrates superior performance, it currently exhibits a larger network size and slower inference speed compared to the state-of-the-art. One potential solution to mitigate this is the adoption of a lightweight model for the LF Encoder and an efficient sampling scheme, as suggested by \cite{song2020denoising}. Another promising direction is to explore LFSRDiff in more LF image processing applications, such as angular SR and depth estimation.

{
    \small
    \bibliographystyle{unsrt} 
    \bibliography{main}
}


\clearpage
\maketitlesupplementary

\section{Appendix}
\label{sec: appendix}

The training and inference procedures of LFSRDiff are described in Sec.~\ref{sup: procedure}. The 2$\times$ SR and 4$\times$ SR benchmark results are further quantitatively compared in Sec.~\ref{sup: compare}. Sec.~\ref{sup: analysis}  provides more analysis of the model. Sec.~\ref{sup: visual} presents additional visual results obtained using various methods. 


\subsection{Training and Inference}
\label{sup: procedure}

\noindent \textbf{Training.} The training procedure of LFSRDiff is illustrated in Alg.~\ref{alg: training}. During the training loop, we randomly sample an LR and HR LF image pair $(\mathcal{L}_{LR}, \mathcal{L}_{HR})$ from the dataset $D$ (Line 2). Then, we generate the upsampled LF LR image $up(\mathcal{L}_{LR})$ by the bicubic interpolation method (Line 3). Next, we compute the residual image $\mathcal{L}_{Res}$ by subtracting the upsampled image $up(\mathcal{L}_{LR})$ from the HR $\mathcal{L}_{HR}$ (Line 4). The encoded feature $f_{LF}$ is extracted from a pre-trained LF Encoder $f_{enc}$ (Line 5). The noise $\epsilon$ is randomly sampled from a Gaussian distribution $\mathcal{N}(0,\mathbf{I})$, while the timestep $t$ is sampled from a uniform distribution $\mathrm{Uniform}(1,\cdots ,T)$ (Line 6). Finally, we calculate the loss according to the formula in Line 7 and update the Distg U-Net $f_\theta$ parameters through gradient descent until the model converges.

\begin{algorithm}[htb]
\caption{The training procedures of LFSRDiff.}
\label{alg: training}
\begin{algorithmic}[1]
\REQUIRE  
    $f_{\theta}$: Distg U-Net, $f_{enc}$: pre-trained LF Encoder, LR and HR LF image pairs $D = (\mathcal{L}_{LR}^i, \mathcal{L}_{HR}^i)_{i=1}^N$, $\alpha_{1:T}$: Noise schedule, $T$: diffusion steps.
    \WHILE {not converged}
        \STATE $(\mathcal{L}_{LR}, \mathcal{L}_{HR})\sim D$, random sample;
        \STATE $up(\mathcal{L}_{LR})=\mathrm{bicubic}(\mathcal{L}_{LR})$, upsample LR image;
        \STATE  $\mathcal{L}_{Res}=\mathcal{L}_{HR}-up(\mathcal{L}_{LR})$, compute residual;
        \STATE $f_{LF}=f_{enc}(\mathcal{L}_{LR})$, encode LF LR image;
        \STATE $\epsilon \sim \mathcal{N}(0,\mathbf{I}) $, and $t \sim \mathrm{Uniform}(1,\cdots ,T)$, sample noise and timestamp;
        \STATE $\nabla _{\theta}\left \| \epsilon - f_\theta (\sqrt {\overline \alpha _t}\mathcal{L}_{Res}+\sqrt{(1-\overline{\alpha}_t)}\epsilon,t,f_{LF})\right \|_1$, perform gradient descent;
    \ENDWHILE
\end{algorithmic}
\end{algorithm}

\noindent \textbf{Inference.} We show the inference procedure in Alg.~\ref{alg: inference}. Similar to the training stage, the LR LF image is upsampled by the bicubic interpolation method (Line 1). We encode the LR LF image to obtain the feature $f_{LF}$ (Line 2). Then, we sample noise from a Gaussian distribution $\mathcal{N}(0,\mathbf{I})$ as initial residual $\mathcal{L}_{Res}^T$ (Line 3). Next, we perform a loop from $T$ to 1 to iteratively generate $\mathcal{L}_{Res}^t$ (Lines 4-7). Specifically, $\mathcal{L}_{Res}^t$ can be computed by the formula in Line 6. Finally, the final SR LF image $\mathcal{L}_{SR}$ can be obtained by adding upsampled LR image $\mathcal{L}_{LR}$ to the final residual image $\mathcal{L}_{Res}^0$ (Line 8).  

\begin{algorithm}[tb]
\caption{The inference procedures of LFSRDiff.}
\label{alg: inference}
\begin{algorithmic}[1]
\REQUIRE  
    $f_{\theta}$: Distg U-Net, $f_{enc}$: pre-trained LF Encoder, LR LF image $\mathcal{L}_{LR}$, $\alpha_{1:T}$: Noise schedule, $T$: diffusion steps.
    \STATE $up(\mathcal{L}_{LR})=\mathrm{bicubic}(\mathcal{L}_{LR})$, upsample LR LF image;
    \STATE $f_{LF}=f_{enc}(\mathcal{L}_{LR})$, encode LF LR image;
    \STATE $\mathcal{L}_{Res}^T \sim \mathcal{N}(0,\mathbf{I})$, sample initial residual;
    \FOR{$t=T, \cdots, 1$}
        \STATE $\epsilon \sim \mathcal{N}(0,\mathbf{I}) $, sample noise;
        \STATE $\mathcal{L}_{Res}^{t-1}=\frac{1}{\sqrt{\alpha _t} }(\mathcal{L}_{Res}^{t}-\frac{1-\alpha _t}{\sqrt[]{1-\overline{\alpha}_t } } f_\theta (\mathcal{L}_{Res}^{t},t,f_{LF}))+\sigma _t(\mathcal{L}_{Res}^{t},t)\epsilon $, compute $\mathcal{L}_{Res}^{t-1}$.
    \ENDFOR
\RETURN $\mathcal{L}_{SR} = up(\mathcal{L}_{LR})+\mathcal{L}_{Res}^{0}$, return the final reconstruction. 
\end{algorithmic}
\end{algorithm}

\begin{figure}[t]
  \centering
   \includegraphics[width=\linewidth]{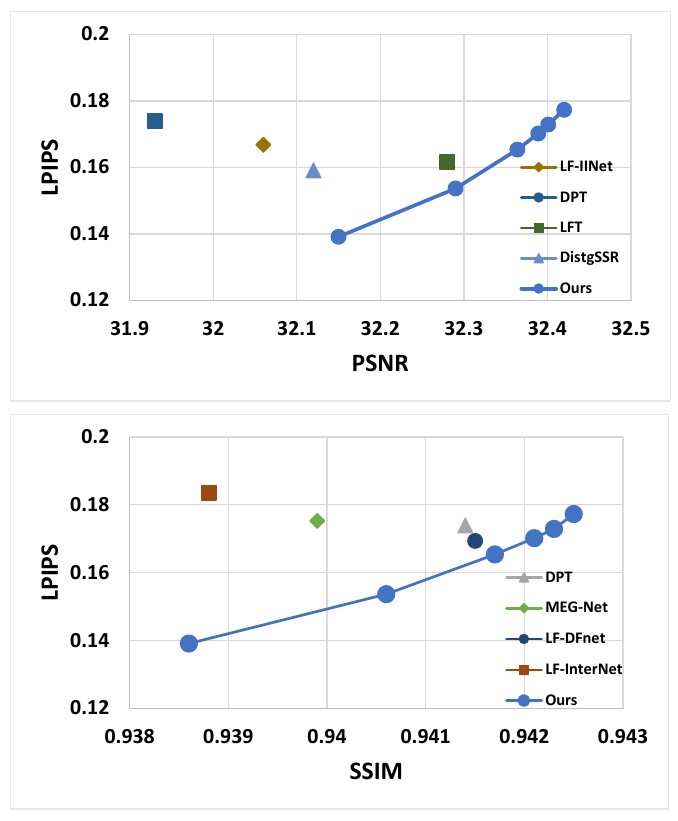}
   \caption{Perception-Distortion plots in terms of different metrics. The upper plot illustrates LPIPS vs. PSNR, and the bottom plot shows SSIM comparisons. Notice that the same trade-off is present for all (perceptual, distortion) metric pairs. }
   \label{sup_fig: p_d_plots}
\end{figure}

\begin{table*}[ht]
\centering
\caption{2$\times$ SR and 4$\times$ SR benchmark results of different SR methods in terms of distortion (PSNR $\uparrow$ and SSIM $\uparrow$) and perception (LPIPS $\downarrow$) metrics. The \textbf{best} results are in bold faces and the {\ul second-best} results are underlined. The right-most column shows the average metrics. }
\renewcommand\arraystretch{1.0}
\resizebox{2 \columnwidth}{!}{
\begin{tabular}{l|ccccc|c}
\toprule
\multirow{2}{*}{\textbf{Methods}} & \multicolumn{5}{|c|}{2$\times$} &  \multirow{2}{*}{\textbf{PSNR/SSIM/LPIPS}}  \\ \cline{2-6} 
& \textbf{EPFL} & \textbf{HCInew} & \textbf{HCIold} & \textbf{INRIA} & \textbf{STFgantry} \\ \midrule
 Bicubic & 29.74/.9376/.1963 & 31.89/.9356/.2241 & 37.69/.9785/.1218 & 31.33/.9577/.1935 & 31.06/.9498/.1553 & 32.34/.9518/.1782 \\
 VDSR \cite{kim2016accurate} (2016) & 32.50/.9598/.0906 & 34.37/.9561/.1215 & 40.61/.9867/.0528  & 34.43/.9741/.0801 & 35.54/.9789/.0346 & 35.49/.9711/.0759 \\
 EDSR \cite{lim2017enhanced} (2017) & 33.09/.9629/.0763 & 34.83/.9592/.1046 & 41.01/.9874/.0495 & 34.97/.9764/.0665 & 36.29/.9818/.0227  & 36.04/.9735/.0639  \\
 RCAN \cite{zhang2018image} (2018)  & 33.16/.9634/.0766 & 34.98/.9603/.1023 & 41.05/.9875/.0490 & 35.01/.9769/.0663 & 36.33/.9831/.0197 & 36.20/.9742/.0628 \\
 SRDiff\cite{li2022srdiff} (2022) & 31.56/.9477/.0588 & 33.21/.9474/.0876 & 39.13/.9832/.0400 & 33.44/.9674/.0581 & 33.86/.9700/.0285 & 34.24/.9631/.0546  \\
 \midrule
 resLF \cite{zhang2019residual} (2019)  & 33.62/.9706/.0540 & 36.69/.9739/.0443 & 43.42/.9932/.0218 & 35.39/.9804/.0548 & 38.36/.9904/.0165 & 37.49/.9817/.0383 \\
 LFSSR \cite{yeung2018light} (2018)  & 33.68/.9744/.0438 & 36.81/.9749/.0447 & 43.81/.9938/.0172 & 35.28/.9832/.0484 & 37.95/.9898/.0211 & 37.50/.9832/.0350 \\
 LF-ATO \cite{jin2020light} (2020)  & 34.27/.9757/.0403 & 37.24/.9767/.0376 & 44.20/.9942/.0172 & 36.15/.9842/.0442 & 39.64/.9929/.0104 & 38.31/.9847/.0299 \\
 LF-InterNet \cite{wang2020spatial} (2020) & 34.14/.9760/.0404 & 37.28/.9763/.0425 & 44.45/.9946/.0149 & 35.80/.9843/.0445 & 38.72/.9909/.0171 & 38.02/.9844/.0319 \\
 LF-DFnet \cite{wang2020light} (2020)  & 34.44/.9755/.0461 & 37.44/.9773/.0565 & 44.23/.9941/.0251 & 36.36/.9840/.0531 & 39.61/.9926/.0391 & 38.39/.9847/.0440 \\
 MEG-Net \cite{zhang2021end} (2021) & 34.30/.9773/.0358 & 37.42/.9777/.0327 & 44.08/.9942/.0150 & 36.09/.9849/.0403 & 38.77/.9915/.0158 & 38.14/.9851/.0279 \\
 LF-IINet \cite{liu2021intra} (2021) & 34.68/.9773/.0340 & 37.74/.9790/.0288 & 44.84/.9948/.0135 & 36.57/.9853/.0388 & 39.86/.9936/.0097 &  38.76/.9860/.0249 \\
 DPT \cite{wang2022detail} (2022) & 34.48/.9758/.0386 & 37.35/.9771/.0358 & 44.31/.9943/.0155 & 36.40/.9843/.0409 & 39.52/.9926/.0106 & 38.40/.9848/.0283 \\
 LFT \cite{liang2022light} (2022) & 34.80/.9781/.0315 & 37.84/.9791/.0284 & 44.52/.9945/.0139 & 36.59/.9855/.0365 & 40.51/.9941/.0075 & 38.85/.9863/.0236 \\
 DistgSSR \cite{wang2022disentangling} (2022) & 34.81/.9787/.0314 & 37.96/.9796/.0264 & 44.94/.9949/.0126 & 36.59/.9859/.0369 & 40.40/.9942/.0072 & 38.94/{\ul .9866}/ {\ul .0229} \\
 EPIT \cite{liang2023learning} (2023) & 34.83/.9775/.0357 & 38.23/.9810/.0228 & 45.08/.9949/.0154  & 36.67/.9853/.0408 & 42.17/.9957/.0045 & \textbf{39.39}/\textbf{.9869}/.0238 \\
 \midrule
 LFSRDiff & 34.00/.9703/.0280 & 37.23/.9761/.0191 & 43.79/.9937/.0140 & 35.86/.9803/.0249 & 40.90/.9945/.0051 & 38.36/.9830/\textbf{.0182} \\
 LFSRDiff (SA) & 34.61/.9759/.0401 & 37.93/.9802/.0253 & 44.37/.9945/.0169 & 36.42/.9842/.0410 & 41.61/.9954/.0056 & {\ul 38.99}/.9860/.0258 \\
\bottomrule
\toprule
\multirow{2}{*}{\textbf{Methods}} & \multicolumn{5}{|c|}{4$\times$} &  \multirow{2}{*}{\textbf{PSNR/SSIM/LPIPS}}  \\ \cline{2-6} 
& \textbf{EPFL} & \textbf{HCInew} & \textbf{HCIold} & \textbf{INRIA} & \textbf{STFgantry}  \\ 
\midrule
 Bicubic & 25.14/.8324/.4617 & 27.61/.8517/.4983 & 32.42/.9344/.3583 & 26.82/.8867/.4368 & 25.93/.8452/.4633 & 27.58/.8701/.4437 \\
 VDSR \cite{kim2016accurate} (2016) & 27.25/.8777/.2795 & 29.31/.8823/.3191 & 34.81/.9515/.1990 & 29.19/.9204/.2611 & 28.51/.9009/.1915 &  29.81/.9066/.2501 \\
 EDSR \cite{lim2017enhanced} (2017) & 27.84/.8854/.2606 & 29.60/.8869/.3117 & 35.18/.9536/.1977 & 29.66/.9257/.2497 & 28.70/.9072/.1671 &  30.19/.9118/.2374 \\
 RCAN \cite{zhang2018image} (2018) & 27.88/.8863/.2594 & 29.63/.8886/.3088 & 35.20/.9548/.2007 & 29.76/.9276/.2471 & 28.90/.9131/.1548 & 30.36/.9141/.2342 \\
 SRDiff\cite{li2022srdiff} (2022) & 26.94/.8655/.2174 & 28.73/.8710/.2702 & 34.16/.9440/.1590 & 28.93/.9118/.2030 & 27.45/.8873/.1454 & 29.24/.8959/.1990  \\
 \midrule
 resLF \cite{zhang2019residual} (2019) & 28.27/.9035/.2351 & 30.73/.9107/.2556 & 36.71/.9682/.1380 & 30.34/.9412/.2244 & 30.19/.9372/.1348 & 31.24/.9322/.1976 \\
 LFSSR \cite{yeung2018light} (2018)  & 28.27/.9118/.2123 & 30.72/.9145/.2458 & 36.70/.9696/.1331 & 30.31/.9467/.2078 & 30.15/.9426/.1235 & 31.52/.9370/.1845 \\
 LF-ATO \cite{jin2020light} (2020) & 28.52/.9115/.2117 & 30.88/.9135/.2534 & 37.00/.9699/.1343 & 30.71/.9484/.2035 & 30.61/.9430/.1212 & 31.54/.9373/.1848 \\
 LF-InterNet \cite{wang2020spatial} (2020) & 28.67/.9162/.2106 & 30.98/.9161/.2444 & 37.11/.9716/.1200 & 30.64/.9491/.2083 & 30.53/.9409/.1341 & 31.61/.9388/.1835 \\
 LF-DFnet \cite{wang2020light} (2020) & 28.77/.9165/.1948 & 31.23/.9196/.2356 & 37.32/.9718/.1213 & 30.83/.9503/.1922 & 31.15/.9494/.1031  & 31.86/.9415/.1694 \\
 MEG-Net \cite{zhang2021end} (2021) & 28.74/.9160/.2076 & 31.10/.9177/.2307 & 37.28/.9716/.1197 & 30.66/.9490/.2021 & 30.77/.9453/.1162  & 31.72/.9399/.1753 \\
 LF-IINet \cite{liu2021intra} (2021) & 29.11/.9188/.2020 & 31.36/.9208/.2274 & 37.62/.9734/.1126 & 31.08/.9515/.1943 & 31.21/.9502/.0978 & 32.06/.9429/.1668 \\
 DPT \cite{wang2022detail} (2022) & 28.93/.9170/.2073 & 31.19/.9188/.2389 & 37.39/.9721/.1216 & 30.96/.9503/.1983 & 31.14/.9488/.1032 & 31.93/.9414/.1739 \\
 LFT \cite{liang2022light} (2022) & 29.25/.9210/.1926 & 31.46/.9218/.2260 & 37.63/.9735/.1074 & 31.20/.9524/.1889 & 31.86/.9548/.0928 & 32.28/{\ul .9447}/.1615 \\
 DistgSSR \cite{wang2022disentangling} (2022) & 28.99/.9195/.1965 & 31.38/.9217/.2203 & 37.56/.9732/.1061 & 30.99/.9519/.1883 & 31.65/.9535/.0843 & 32.12/.9440/{\ul .1591} \\
 EPIT \cite{liang2023learning} (2023) & \textbf 29.34/.9197/.1999 & 31.51/.9231/.2206 & 37.68/.9737/.1140 & 31.27/.9526/.1893 & 32.18/.9571/.0842 & {\ul 32.41}/\textbf{.9452}/.1616  \\
 \midrule
 LFSRDiff & 29.38/.9107/.1645 & 31.10/.9149/.1819 & 37.13/.9704/.1031 & 31.58/.9471/.1535 & 31.58/.9502/.0929 & 32.15/.9386/\textbf{.1392} \\
 LFSRDiff (SA) & 29.62/.9164/.2141 & 31.32/.9199/.2366 & 37.45/.9727/.1256 & 31.86/.9502/.2027 & 31.85/.9534/.1077 & \textbf{32.42}/.9425/.1773 \\
\bottomrule
\end{tabular}
}

\label{table: quantitative2}
\end{table*}

\begin{figure}[htb]
  \centering
   \includegraphics[width=\linewidth]{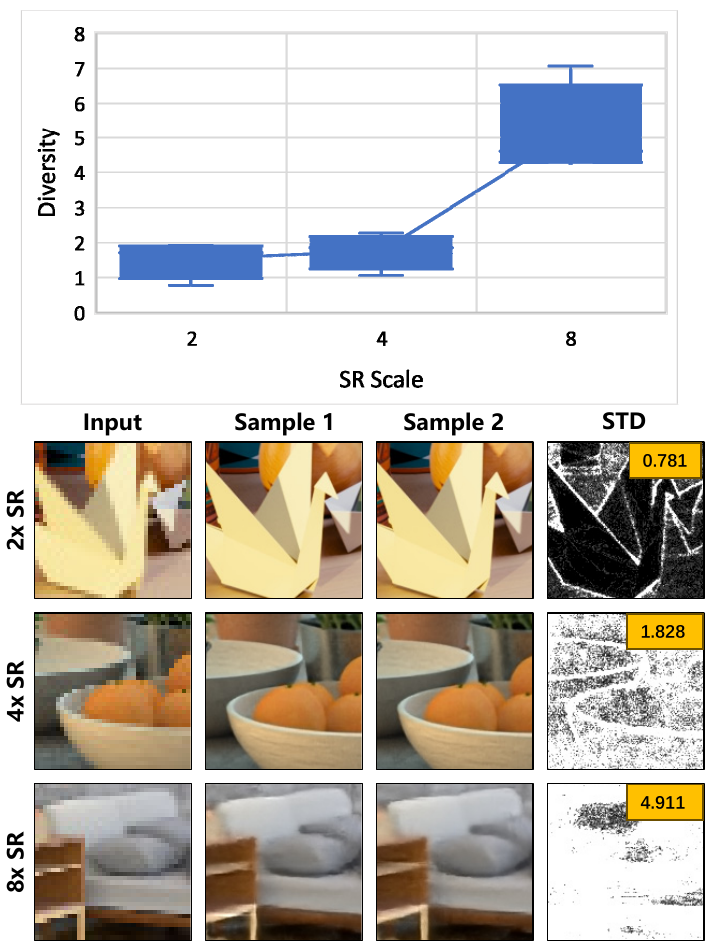}
   \caption{Sample diversity in terms of SR scale factor. The upper plot illustrates the Diversity vs. SR scale. The bottom shows visual comparisons. The ill-posed SR task (i.e., what the SR scale factor is) has a positive impact on the diversity of the generated samples. The more the white regions, the larger the diversity. }
   \label{sup_fig: diversity}
\end{figure}

\begin{figure}[htb]
  \centering
   \includegraphics[width=\linewidth]{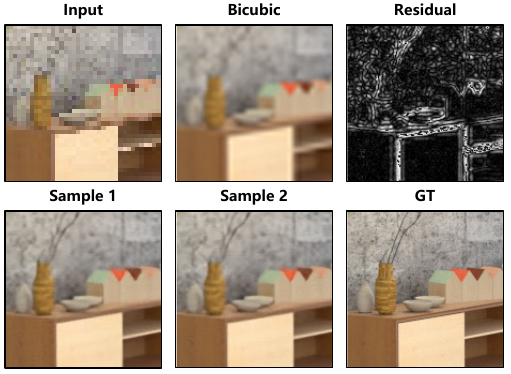}
   \caption{Output of the bicubic and multiple samples generated from our LFSRDiff. We see that the over-smoothed bicubic upsampling lacking texture is corrected by the LFSRDiff, producing realistic and diverse final reconstructions. The top right figure shows the residual predicted by our LFSRDiff.}
   \label{sup_fig: residual}
\end{figure}

\subsection{Benchmark Comparisons}
\label{sup: compare}

Table~\ref{table: quantitative2} displays the distortion (PSNR and SSIM) and perception (LPIPS) metrics of our method compared to state-of-the-art methods on the 2$\times$ SR and 4$\times$ SR tasks. Since our model is supervised by predicting noise rather than pixel-wise loss,
it can generate diverse results and achieve the best perceptual metric on both the 2$\times$ SR and 4$\times$ SR tasks, which are more consistent with human perception. By employing the sample averaging (SA) strategy, our model achieves the best PSNR on the 4$\times$ SR and the second-best PSNR on the 2$\times$ SR. Note that, theoretically, we can achieve better PSNR metrics by employing self-ensemble with an infinite number of results. However, in this paper, we only report the results obtained using 16 samples. Therefore, the sample averaging strategy can be adjusted to achieve a favorable trade-off between perception and distortion, depending on specific requirements.

\begin{figure}[htb]
  \centering
   \includegraphics[width=\linewidth]{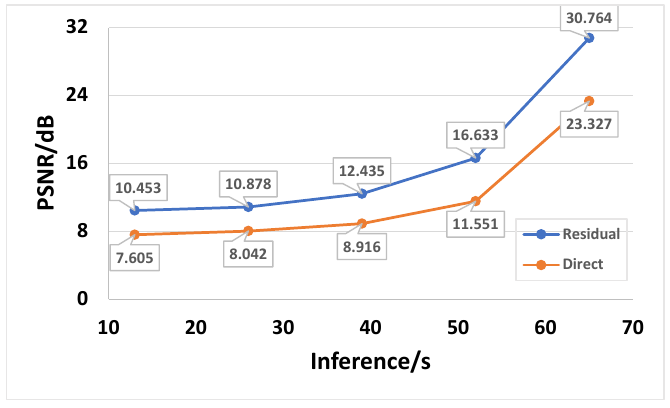}
   \caption{Plot of PSNR vs. Inference Time. The residual modeling achieves lower latency while maintaining higher sample quality than direct learning.}
   \label{sup_fig: inference}
\end{figure}

\subsection{Model Analysis}
\label{sup: analysis}

\noindent \textbf{Perception-Distortion Plots.} The trade-off between perception (LPIPS) and distortion (PSNR) has been validated in the main text of this paper. We observe that other combinations of perceptual (LPIPS) and distortion metrics (PSNR, SSIM) follow a similar trend, as shown in Fig.~\ref{sup_fig: p_d_plots}. Note that LPIPS is more consistent with human perception than PSNR or SSIM.

\noindent \textbf{Diversity Analysis.} Fig.~\ref{sup_fig: diversity} illustrates the relation between the SR scale factor and the diversity of the generated super-resolved samples. We employ average per-pixel standard deviation (STD) to represent diversity. As the SR scaling factor increases, the diversity of generated samples also increases. We can clearly observe pixel-level differences, such as in the case of 8$\times$ SR, as shown in the bottom part of Fig.~\ref{sup_fig: diversity}.

\noindent \textbf{Residual Modeling.} As shown in Fig.~\ref{sup_fig: residual}, it is evident that the upsampled LR image suffers from noticeable blurriness, lacking fine details. However, our LFSRDiff approach effectively resolves this issue by predicting and incorporating the residuals between the upsampled image and the ground truth (GT) image. Consequently, the over-smoothed upsampled LR image, which previously lacked texture, is rectified by the LFSRDiff model, resulting in more realistic and diverse final reconstructions.



\begin{figure*}[htb]
  \centering
   \includegraphics[width=\linewidth]{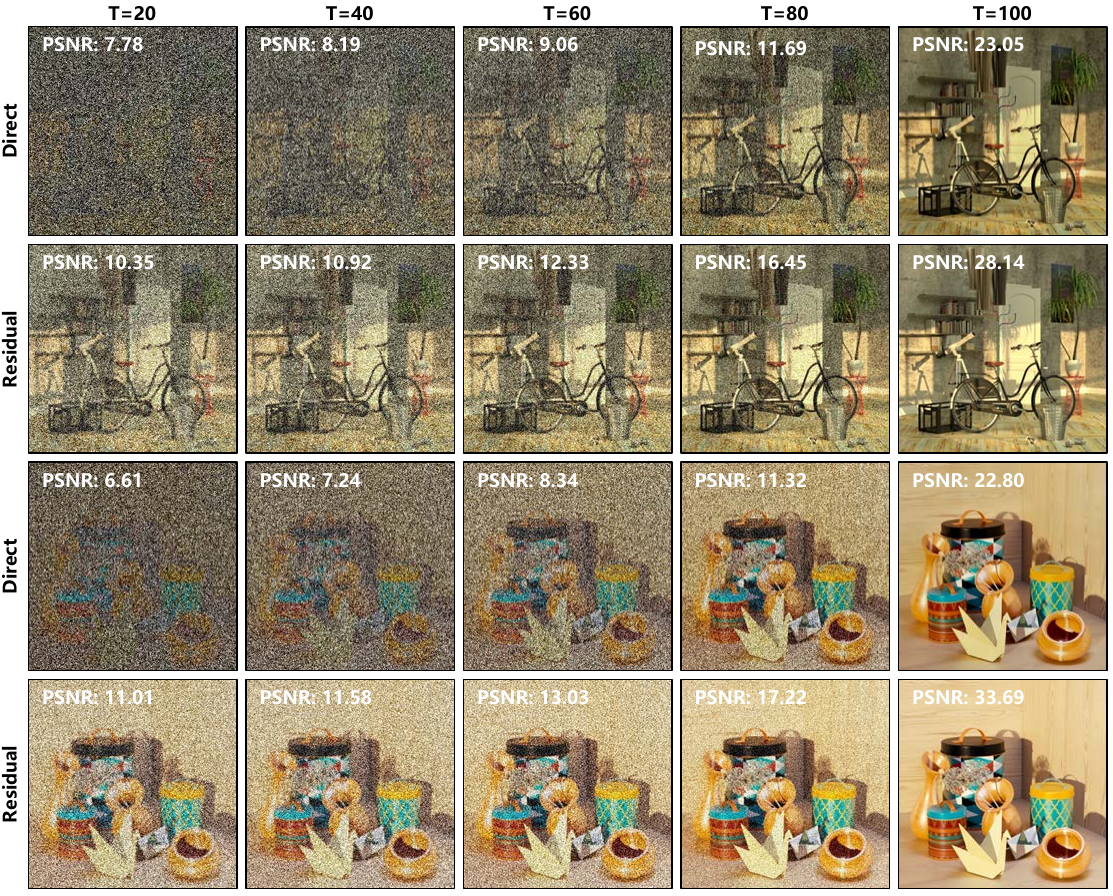}
   \caption{Visualization of intermediate results in direct learning and residual modeling. By examining the visual representations, we aim to gain insights into the differences and advantages of residual modeling over direct learning.}
   \label{sup_fig: intermediate}
\end{figure*}

The primary benefit of residual modeling lies in the reduced computational cost associated with sampling. Due to the iterative process of diffusion sampling, the U-Net requires numerous iterations for each generated sample, typically ranging from several dozen to hundreds of steps. Consequently, any decrease in the computational burden imposed by the U-Net is highly valuable. Our upsampled sample offers a straightforward means to alleviate a portion of this computation. As shown in Fig.~\ref{sup_fig: inference}, residual modeling requires considerably less time to sample an image compared to direct learning and achieves higher PSNR for the same sampling steps. Fig.~\ref{sup_fig: intermediate} visualizes the intermediate results of direct learning and residual modeling. It is evident from the visualization that residual modeling exhibits significantly superior visual perception compared to direct learning, which demonstrates the importance of residual learning.

\begin{figure*}[htb]
  \centering
   \includegraphics[width=\linewidth]{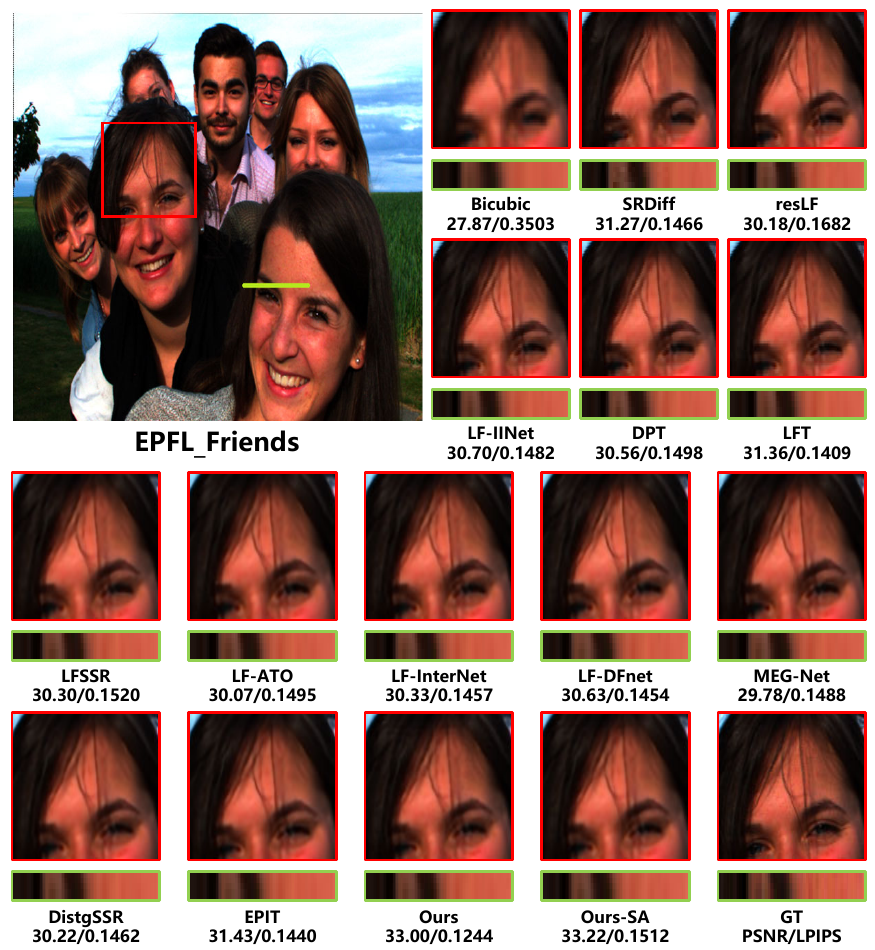}
   \caption{Qualitative comparison of different SR methods for 4$\times$ SR in Scene \textit{Friends} from EPFL dataset \cite{rerabek2016new}. The super-resolved center view images and horizontal EPIs are shown.}
   \label{sup_fig: friends}
\end{figure*}

\begin{figure*}[htb]
  \centering
   \includegraphics[width=\linewidth]{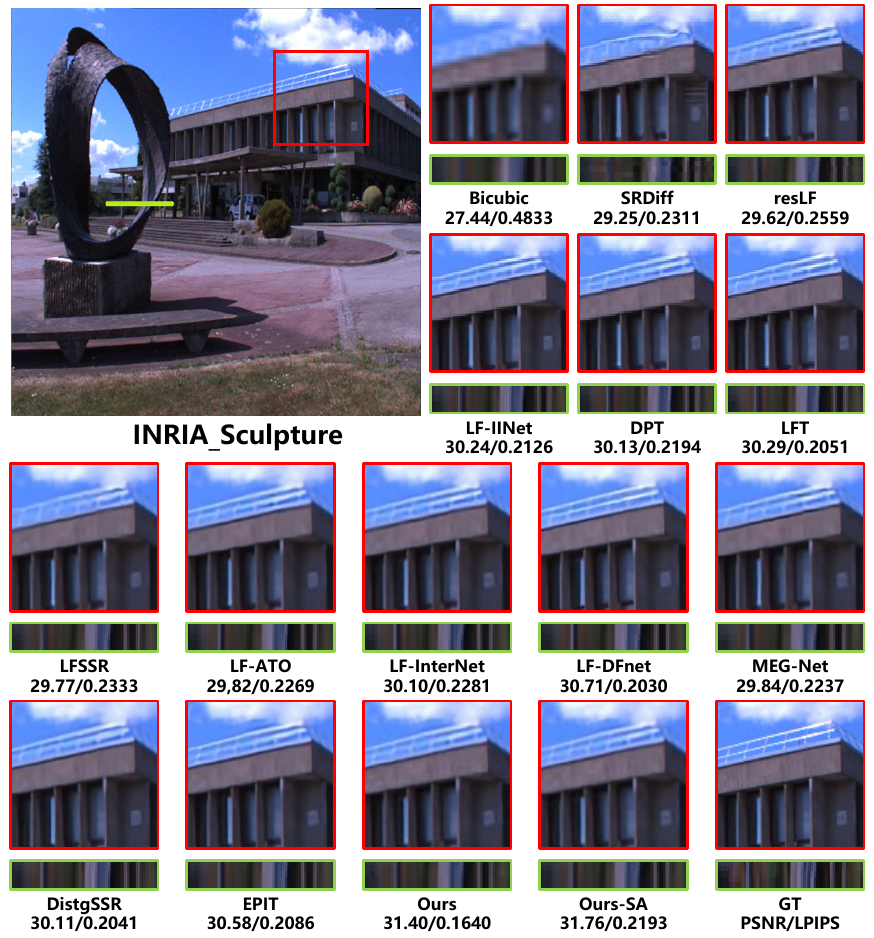}
   \caption{Qualitative comparison of different SR methods for 4$\times$ SR in Scene \textit{Sculpture} from INRIA dataset \cite{le2018light}. The super-resolved center view images and horizontal EPIs are shown.}
   \label{sup_fig: sculpture}
\end{figure*}

\begin{figure*}[htb]
  \centering
   \includegraphics[width=\linewidth]{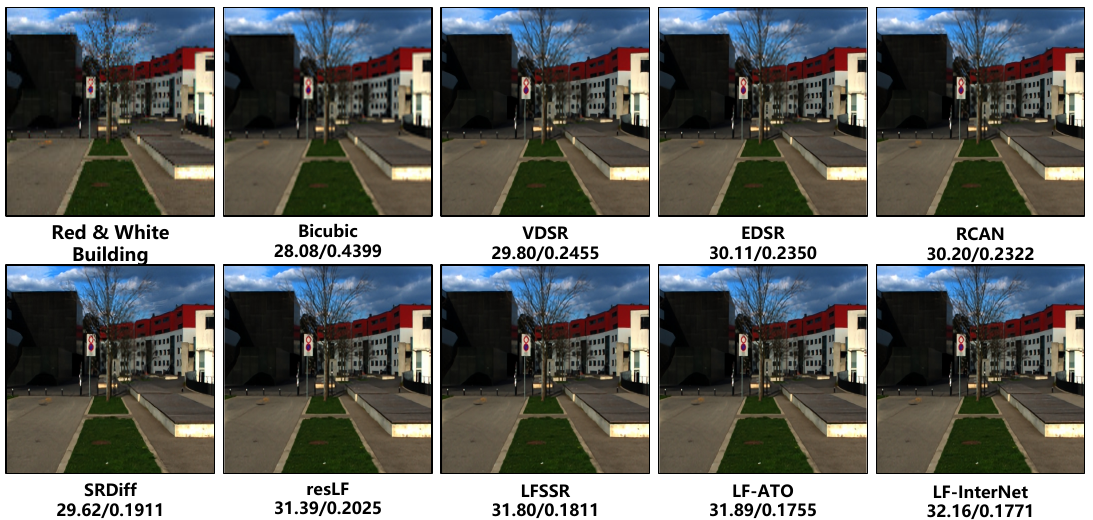}
   \includegraphics[width=\linewidth]{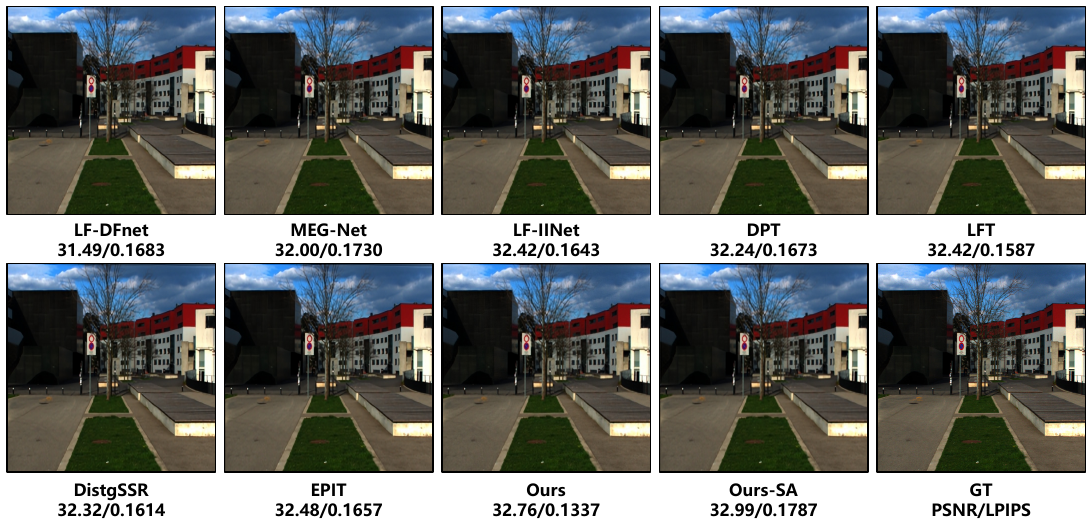}
   \caption{Visual comparison of different SR methods for 4$\times$ SR in real Scene \textit{Buliding} from EPFL \cite{rerabek2016new} dataset.}
   \label{sup_fig: building}
\end{figure*}

\begin{figure*}[htb]
  \centering
   \includegraphics[width=\linewidth]{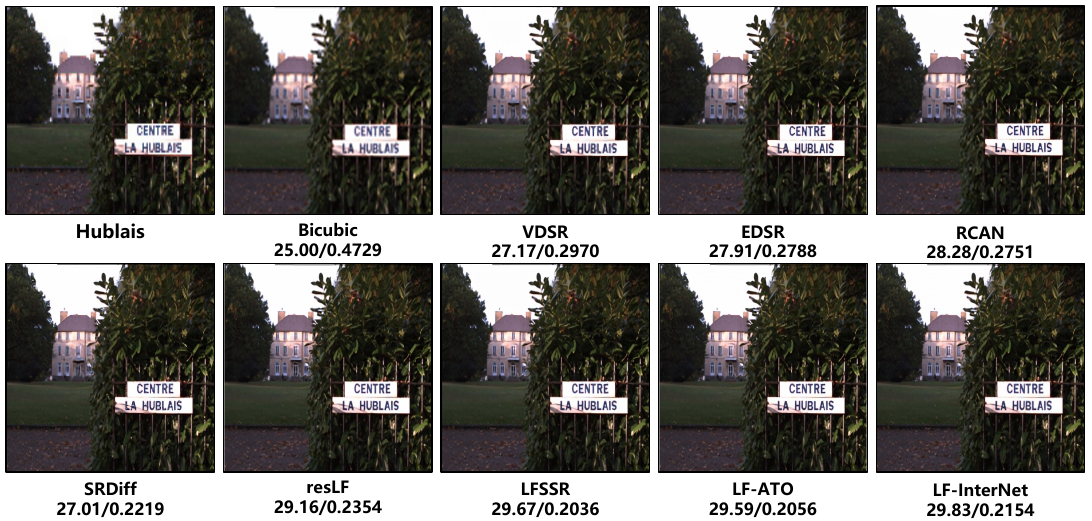}
   \includegraphics[width=\linewidth]{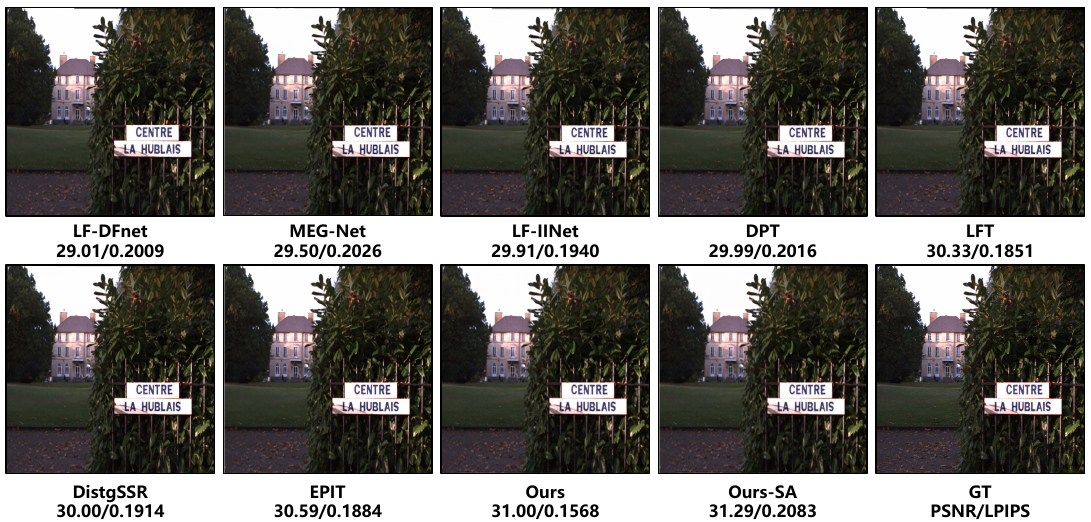}
   \caption{Visual comparison of different SR methods for 4$\times$ SR in real Scene \textit{Sculpture} from INRIA dataset \cite{le2018light}.}
   \label{sup_fig: hublais}
\end{figure*}

\subsection{Additional Visual Comparisons}
\label{sup: visual}

In this subsection, we present additional visual comparisons of 4$\times$ SR on the benchmark datasets. We compare our LFSRDiff to 15 state-of-the-art (SOTA) methods, including 4 single image SR methods \cite{kim2016accurate, lim2017enhanced, zhang2018image, li2022srdiff} and 11 LF image SR methods \cite{zhang2019residual, yeung2018light, jin2020light, wang2020spatial, wang2020light, zhang2021end, liu2021intra, wang2022detail, liang2022light, wang2022disentangling, liang2023learning}. Consistent with the main text, \textit{Ours-SA} refers to the sample-averaging variant of our method. In \cref{sup_fig: friends,sup_fig: sculpture,sup_fig: intermediate}, we include larger versions of the EPFL and INTRA scenes shown in the main text. In \cref{sup_fig: building,sup_fig: hublais}, we present additional results on the real scenes from EPFL \cite{rerabek2016new} and INRIA \cite{le2018light} dataset. It is evident that the proposed LFSRDiff method yields better performance in recovering more intricate and realistic details than the SOTA. 

\end{document}